\documentclass[preprint]{ptephy_om}

\preprintnumber{KYUSHU-HET-336} 

\usepackage{graphics}

\usepackage{amsmath} 
\usepackage{amsthm} 
\usepackage{url} 
\usepackage{stmaryrd}
\usepackage{tikz}
\usepackage{tikz-feynhand}
\usepackage{mathtools}
\usepackage{bm}
\usepackage{booktabs}
\usepackage{subcaption}

\numberwithin{equation}{section}

\allowdisplaybreaks

\begin{document}

\title{Yang--Mills $\beta$ function in the gradient flow exact renormalization
group}


\author[1]{Sorato Nagao}
\affil[1]{Department of Physics, Kyushu University, 744 Motooka, Nishi-ku,
Fukuoka 819-0395, Japan}

\author[1]{Hiroshi Suzuki}





\begin{abstract}%
The gradient flow exact renormalization (GFERG) is a variant of the exact
renormalization group of gauge theory that aims to preserve gauge symmetry
as manifestly as possible. From an integral representation of the Wilson
action in GFERG for the Yang--Mills theory, we explicitly compute the one-loop
renormalization group functions that reproduce correct coefficients. From the
correspondence with the gradient flow formalism by L\"uscher and Weisz, we also
argue that GFERG reproduces the conventional renormalization group functions in
all orders of perturbation theory.
\end{abstract}

\subjectindex{B01,B05,B31,B32}

\maketitle

\section{Introduction}
\label{sec:1}
The momentum cutoff conventionally assumed in the Wilson exact
renormalization group (ERG)~\cite{Wilson:1973jj,Morris:1993qb,Becchi:1996an,Pawlowski:2005xe,Igarashi:2009tj,Rosten:2010vm,Dupuis:2020fhh} is incompatible
with a manifest gauge invariance and, for this reason, it is extremely
difficult to study a non-perturbative ERG flow while keeping track only of
gauge invariant physical contents. The gradient flow exact renormalization
group (GFERG), considered
in~Refs.~\cite{Sonoda:2020vut,Miyakawa:2021hcx,Miyakawa:2021wus,Sonoda:2022fmk,Miyakawa:2023yob,Sonoda:2025tyu}, is a variant of ERG that aims to keep a
manifest gauge or BRST invariance along the ERG flow;
see~Refs.~\cite{Abe:2022smm,Miyakawa:2022qbz,Haruna:2023spq} for related
works.\footnote{Refs.~\cite{Morris:1999px,Morris:2000fs,Arnone:2005fb,Morris:2005tv} are prior researches on a manifestly gauge invariant ERG.}

In a recent paper~\cite{Sonoda:2025tyu}, GFERG was reformulated starting from
the Reuter equation representation of ERG.\footnote{%
See, for instance, Eqs.~(2.4) and~(2.5) of~Ref.~\cite{Reuter:2019byg}.} This
reformulation solved several issues that were not settled in previous
papers~\cite{Sonoda:2020vut,Miyakawa:2021hcx,Miyakawa:2021wus,Sonoda:2022fmk,Miyakawa:2023yob}. One is the ordering of functional derivatives in the GFERG
flow equation; an ordering being perfectly consistent was clarified. Another is
a finiteness issue of GFERG without the gauge fixing. Also, the connection with
the gradient flow formalism
in~Refs.~\cite{Luscher:2010iy,Luscher:2011bx,Luscher:2013cpa} became much more
transparent; the modified correlation functions~\cite{Sonoda:2015bla} given by
the GFERG Wilson action reproduce correlation functions in the gradient flow
formalism up to contact terms. The Reuter equation also provides a very
powerful method to obtain the perturbative solution of the GFERG flow equation.
The GFERG flow equation for the Wilson action is of 4th order in functional
derivatives (see~Eq.~\eqref{eq:(2.9)} below for the case of the pure
Yang--Mills theory), whereas the flow equation in the conventional ERG is of
2nd order. This is the sacrifice required for a manifest gauge or BRST
invariance. For this reason, directly solving the GFERG flow equation is highly
complex even in perturbation theory. An attempt to directly solve the GFERG
flow equation for QED in perturbation theory can be found
in~Ref.~\cite{Miyakawa:2021wus}; one has to solve the differential equation
such as~Eq.~(5.41) there by finding ingenious manipulations. For the
Yang--Mills theory, the required effort becomes even much
higher.\footnote{H.~Sonoda and H.~Suzuki, unpublished work.}

In the present paper, by employing the Reuter equation in GFERG, i.e., a
functional integral representation for the Wilson action, we compute the
one-loop renormalization group functions, such as the $\beta$-function, in the
pure Yang--Mills theory.\footnote{%
For previous computations of the renormalization group functions in the
conventional ERG, see~Refs.~\cite{Bonini:1996bk,Igarashi:2019gkm} and
references cited therein.} The calculational labor is considerably less
compared with a direct solution to the GFERG flow equation, i.e., the
functional differential equation for the Wilson action; what one has to carry
out is simply a routine expansion of the integrand of the functional integral
and no ingenious manipulation is required. The computation reproduces the known
correct results; see~Eq.~\eqref{eq:(2.33)}. We can also argue that, on the
basis of the connection to the gradient flow formalism, GFERG reproduces the
conventional renormalization group functions in all orders of perturbation
theory. The perturbative computation rule turns to be quite similar to that
in~Refs.~\cite{Luscher:2010iy,Luscher:2011bx,Luscher:2013cpa} as expected. We
emphasize, however, that our perturbation theory is the one to compute the
Wilson action in GFERG as a whole. The Feynman rule for the Wilson action is
summarized in~Appendix~\ref{sec:A}.

\section{Yang--Mills $\beta$ function in GFERG}
\subsection{Wilson action in GFERG}
\label{sec:2.1}
As discussed in~Ref.~\cite{Sonoda:2025tyu}, it is convenient to represent the
Wilson action~$S_\Lambda$ in GFERG, where $\Lambda$ is the ultraviolet (UV)
cutoff in the Wilson action, by an integral representation, the Reuter equation
(see also~Refs.~\cite{Miyakawa:2021wus,Miyakawa:2023yob}). For the pure
Yang--Mills theory, it reads\footnote{%
Throughout this paper, $D$ denotes the spacetime dimension and we set
\begin{equation}
   \epsilon:=\frac{4-D}{2}.
\label{eq:(2.1)}
\end{equation}}
\begin{align}
   &e^{S_\Lambda[A,\Bar{c},c]}
\notag\\
   &:=\mathcal{N}(\Lambda)
   \int[dA'][dc'][d\Bar{c}']\,
\notag\\
   &\qquad\qquad{}
   \times\exp\left\{
   -\frac{\Lambda^2}{2}\int d^Dx\,
   \left[
   A_\mu^a(x)-\frac{1}{\Lambda^\epsilon g_\Lambda}A_\mu^{\prime a}(t,x)
   \right]^2\right\}
\notag\\
   &\qquad\qquad{}
   \times\exp\left\{
   -\Lambda^2\int d^Dx\,
   \left[\Bar{c}^a(x)-Z_{\Bar{c}}(\Lambda)
   \frac{1}{\Lambda^\epsilon g_\Lambda}\Bar{c}^{\prime a}(x)\right]
   \left[c^a(x)-\frac{1}{\Lambda^\epsilon g_\Lambda}c^{\prime a}(t,x)\right]
   \right\}
\notag\\
   &\qquad\qquad\qquad{}
   \times e^{S[A',\Bar{c}',c']},
\label{eq:(2.2)}
\end{align}
where $Z_{\Bar{c}}(\Lambda)$ is the wave function renormalization factor for the
Faddeev--Popov (FP) anti-ghost. Compared with~Eq.~(5.1)
of~Ref.~\cite{Sonoda:2025tyu}, in~Eq.~\eqref{eq:(2.2)}, we have introduced the
renormalized (dimensionless) gauge coupling~$g_\Lambda$; Eq.~\eqref{eq:(2.19)}
illustrates that $g_\Lambda$ precisely provides the expansion parameter in
perturbation theory. The normalization factor~$\mathcal{N}(\Lambda)$ is defined
by
\begin{align}
   \mathcal{N}(\Lambda)
   &:=\biggl\{
   \int[dA][dc][d\Bar{c}]\,
\notag\\
   &\qquad\qquad{}
   \times
   \exp\left[
   -\frac{\Lambda^{D-2}}{2}\int d^Dx\,A_\mu^a(x)^2\right]
   \exp\left[
   -\Lambda^{D-2}\int d^Dx\,\Bar{c}^a(x)c^a(x)\right]
   \biggr\}^{-1}.
\label{eq:(2.3)}
\end{align}
As the bare action~$S$ in~Eq.~\eqref{eq:(2.2)}, we adopt
\begin{align}
   &S[A',\Bar{c}',c']
\notag\\
   &=-\frac{1}{g_0^2}\int d^Dx
   \left[
   \frac{1}{4}F_{\mu\nu}^{\prime a}(x)F_{\mu\nu}^{\prime a}(x)
   +\frac{1}{2\xi_0}\partial_\mu A_\mu^{\prime a}(x)
   \partial_\nu A_\nu^{\prime a}(x)\right]
   +\frac{1}{g_0^2}\int d^Dx\,
   \Bar{c}^{\prime a}(x)\partial_\mu D_\mu'c^{\prime a}(x),
\label{eq:(2.4)}
\end{align}
where $g_0$ and~$\xi_0$ are bare parameters.

In GFERG, $A_\mu^{\prime a}(t,x)$ in the integrand of~Eq.~\eqref{eq:(2.2)} is
given by the solution of the Yang--Mills gradient flow
equation~\cite{Luscher:2010iy,Luscher:2011bx}, whose initial
value is the integration variable~$A_\mu^{\prime a}(x)$:
\begin{equation}
   \partial_tA_\mu'(t,x)=D_\nu'F_{\nu\mu}'(t,x)
   +\alpha_0D_\mu'\partial_\nu A_\nu'(t,x),\qquad
   A_\mu'(t=0,x)=A_\mu'(x),
\label{eq:(2.5)}
\end{equation}
where $\alpha_0>0$ is a constant and
\begin{align}
   F_{\mu\nu}'(t,x)&:=\partial_\mu A_\nu'(t,x)-\partial_\nu A_\mu'(t,x)
   +[A_\mu'(t,x),A_\nu'(t,x)],
\notag\\
   D_\mu'&:=\partial_\mu+[A_\mu'(t,x),\phantom{X}].
\label{eq:(2.6)}
\end{align}
Similarly, the flowed or diffused FP ghost, $c^{\prime a}(t,x)$ is given from the
integration variable~$c^{\prime a}(x)$ by~\cite{Luscher:2010iy,Luscher:2011bx}
\begin{equation}
   \partial_tc'(t,x)=\alpha_0D_\mu'\partial_\mu c'(t,x),\qquad
   c'(t=0,x)=c'(x).
\label{eq:(2.7)}
\end{equation}
In this way, the GFERG flow equation preserves the underlying BRST
symmetry (see~Ref.~\cite{Sonoda:2025tyu}). In these expressions, the flow or
diffusion time~$t$ and the cutoff~$\Lambda$ are related by
\begin{equation}
   t:=\frac{1}{\Lambda^2}-\frac{1}{\Lambda_0^2},
\label{eq:(2.8)}
\end{equation}
where $\Lambda_0$ denotes the cutoff that defines the bare theory. Actually,
in the present paper, we employ the dimensional regularization
with~$\epsilon=(4-D)/2$ to define the bare theory and we identify the
``continuum limit''~$\Lambda_0\to\infty$ with the limit~$\varepsilon\to0$.

The GFERG flow equation for the Wilson action~$S_\Lambda$ can be obtained simply
by taking the $\Lambda$~derivative
of~Eq.~\eqref{eq:(2.2)}~\cite{Sonoda:2025tyu}:
\begin{align}
   &-\Lambda\frac{\partial}{\partial\Lambda}
   e^{S_\Lambda[A,\Bar{c},c]}
\notag\\
   &=\int d^Dx\,
   \biggl(
   \frac{\delta}{\delta A_\mu^a(x)}
   \biggl\{
   -\frac{2}{\Lambda^2}
   \left[
   \Hat{D}_\nu \Hat{F}_{\nu\mu}^a(x)
   +\alpha_0\Hat{D}_\mu\partial_\nu\Hat{A}_\nu^a(x)
   \right]
   +\left(-\epsilon-\Lambda\frac{d}{d\Lambda}\ln g_\Lambda\right)\Hat{A}_\nu^a(x)
\notag\\
   &\qquad\qquad\qquad\qquad\qquad{}
   +\frac{1}{\Lambda^2}\frac{\delta}{\delta A_\mu^a(x)}
   \biggr\}
\notag\\
   &\qquad\qquad\qquad{}
   +\frac{\delta}{\delta c^a(x)}
   \left[
   \frac{2}{\Lambda^2}
   \alpha_0\Hat{D}_\mu\partial_\mu\Hat{c}^a(x)
   +\left(\epsilon+\Lambda\frac{d}{d\Lambda}\ln g_\Lambda\right)\Hat{c}^a(x)
   -\frac{2}{\Lambda^2}
   \frac{\delta}{\delta\Bar{c}^a(x)}\right]
\notag\\
   &\qquad\qquad\qquad{}
   +\frac{\delta}{\delta\Bar{c}^a(x)}
   \left[-\Lambda\frac{d}{d\Lambda}\ln Z_{\Bar{c}}(\Lambda)
   +\epsilon+\Lambda\frac{d}{d\Lambda}\ln g_\Lambda\right]
   \Hat{\Bar{c}}^a(x)
   \biggr)\,
   e^{S_\Lambda[A,\Bar{c},c]},
\label{eq:(2.9)}
\end{align}
where variables with the hat~($\Hat{\phantom{X}}$) denote differential
operators defined by
\begin{align}
   \Hat{A}_\mu^a(x)
   &:=A_\mu^a(x)+\frac{1}{\Lambda^2}\frac{\delta}{\delta A_\mu^a(x)},&&
\notag\\
   \Hat{c}^a(x)&:=c^a(x)+\frac{1}{\Lambda^2}
   \frac{\delta}{\delta\Bar{c}^a(x)},&
   \Hat{\Bar{c}}^a(x)&:=
   \Bar{c}^a(x)-\frac{1}{\Lambda^2}\frac{\delta}{\delta c^a(x)}.
\label{eq:(2.10)}
\end{align}
Multiple products of hatted field variables accompany additional factors
of~$\Lambda^\epsilon g_\Lambda$; this shows that $g_\Lambda$ provides the coupling
constant. In~Eq.~\eqref{eq:(2.9)}, the ordering of functional derivatives is
carefully chosen so that no additional contact terms
arise~\cite{Sonoda:2025tyu}. Under this GFERG flow equation, the Wilson action
fulfills the Ward--Takahashi (WT) identity associated with the BRST invariance
of the bare action~$S$~\cite{Sonoda:2025tyu},\footnote{%
The structure constants~$f^{abc}$ are defined from (anti-Hermitian) gauge group
generators~$T^a$ by~$[T^a,T^b]=f^{abc}T^c$. The Dynkin index~$C_A$ is defined
by~$f^{acd}f^{bcd}=C_A\delta^{ab}$.}
\begin{align}
   &\int d^Dx\,
   \biggl\{
   \frac{\delta}{\delta A_\mu^a(x)}\Hat{D}_\mu\Hat{c}^a(x)
   -\Lambda^\epsilon g_\Lambda
   \frac{\delta}{\delta c^a(x)}\frac{1}{2}f^{abc}\Hat{c}^b(x)\Hat{c}^c(x)
\notag\\
   &\qquad\qquad{}
   +\frac{1}{\Lambda^\epsilon g_\Lambda}\frac{\delta}{\delta\Bar{c}^a(x)}
   \frac{Z_{\Bar{c}}(\Lambda)}{\xi_0}\partial_\mu
   \mathcal{I}_\mu^a\left[\Hat{A};(t,x)\right]
   \biggr\}\,e^{S_\Lambda[A,\Bar{c},c]}=0.
\label{eq:(2.11)}
\end{align}
In this expression, $\mathcal{I}_\mu^a[\Hat{A};(t,x)]$ denotes the initial
condition of~Eq.~\eqref{eq:(2.5)} which leads~$A_\mu(t,x)$ as the
solution~\cite{Sonoda:2025tyu}. More specifically,
$\mathcal{I}_\mu^a[\Hat{A};(t,x)]$ is the initial condition of the gradient flow
equation such that
\begin{equation}
   \partial_sA_\mu(s,x)=D_\nu F_{\nu\mu}(s,x)
   +\alpha_0D_\mu\partial_\nu A_\nu(s,x),\qquad
   A_\mu(s=0,x)=\mathcal{I}_\mu\left[\Hat{A};(t,x)\right]
\label{eq:(2.12)}
\end{equation}
and $A_\mu(s=t,x)=A_\mu(t,x)$.

\subsection{Renormalization group functions}
\label{sec:2.2}
In what follows, we set renormalization constants as
\begin{equation}
   g_0^2=\Lambda^{2\epsilon}g_\Lambda^2 Z,\qquad
   \xi_0=\xi_\Lambda Z_3
\label{eq:(2.13)}
\end{equation}
and determine the one loop coefficients $b_0$ and~$c_0$ in
\begin{equation}
   Z=1-\frac{b_0}{\epsilon}g_\Lambda^2+O(g_\Lambda^4),\qquad
   Z_3=1+\frac{c_0}{\epsilon}g_\Lambda^2+O(g_\Lambda^4),
\label{eq:(2.14)}
\end{equation}
by requiring that the Wilson action~$S_\Lambda$ remains finite as~$\epsilon\to0$.

It should be possible to solve the flow equation~\eqref{eq:(2.9)} in the
perturbative series in~$g_\Lambda$ with an appropriate boundary conditions that
is consistent with the WT identity~\eqref{eq:(2.11)}. However, if we assume an
explicit form of the bare action as~Eq.~\eqref{eq:(2.4)}, the direct
perturbative expansion of the integral~\eqref{eq:(2.2)} is much simpler; this
is our strategy in the present paper.

Now, setting the gauge kinetic term in the Wilson action as\footnote{%
We use the abbreviation~$\int_p:=\int d^Dp/(2\pi)^D$.}
\begin{equation}
   S_\Lambda
   =-\frac{1}{2}\int_p A_\mu^a(-p)A_\nu^b(p)
   \mathcal{K}_\Lambda(p)_{\mu\nu}^{ab}+\dotsb.
\label{eq:(2.15)}
\end{equation}
Eq.~\eqref{eq:(2.2)} shows that the kernel~$\mathcal{K}_\Lambda(p)_{\mu\nu}^{ab}$ is
given by the two-point function,
\begin{equation}
   \left\langle A_\mu^{\prime a}(t,p)A_\nu^{\prime b}(t,q)\right\rangle
   =:(2\pi)^D\delta(p+q)G_{2t}(p)_{\mu\nu}^{ab},
\label{eq:(2.16)}
\end{equation}
as
\begin{equation}
   \mathcal{K}_\Lambda(p)_{\mu\nu}^{ab}
   =\Lambda^2\delta^{ab}\delta_{\mu\nu}
   -\frac{\Lambda^4}{\Lambda^{2\epsilon}g_\Lambda^2}
   G_{2t}(p)_{\mu\nu}^{ab}.
\label{eq:(2.17)}
\end{equation}

The tree-level two-point function given in~Eqs.~\eqref{eq:(A20)}
and~\eqref{eq:(A21)} contributes to this by
\begin{align}
   &\Lambda^2\delta^{ab}\delta_{\mu\nu}
   -\frac{\Lambda^4}{\Lambda^{2\epsilon}g_\Lambda^2}
   G_{2t}^{\text{tree}}(p)_{\mu\nu}^{ab}
\notag\\
   &=
   \delta^{ab}\left[\Lambda^2\delta_{\mu\nu}
   -\Lambda^4ZD_{2t}(p)_{\mu\nu}\right]
\notag\\
   &=\delta^{ab}\left[
   (\delta_{\mu\nu}p^2-p_\mu p_\nu)\frac{1}{p^2/\Lambda^2+e^{-2tp^2}}
   +p_\mu p_\nu\frac{1}{p^2/\Lambda^2+\xi_\Lambda e^{-2\alpha_0tp^2}}\right]
\notag\\
   &\qquad{}
   +g_\Lambda^2\delta^{ab}\left[
   \frac{b_0}{\epsilon}
   (\delta_{\mu\nu}p^2-p_\mu p_\nu)\frac{e^{-2tp^2}}{(p^2/\Lambda^2+e^{-2tp^2})^2}
   +\frac{b_0-c_0}{\epsilon}
   p_\mu p_\nu\frac{\xi_\Lambda e^{-2\alpha_0tp^2}}
   {(p^2/\Lambda^2+\xi_\Lambda e^{-2\alpha_0tp^2})^2}\right]
\notag\\
   &\qquad{}+O(g_\Lambda^4),
\label{eq:(2.18)}
\end{align}
where we have used~Eqs.~\eqref{eq:(2.13)} and~\eqref{eq:(2.14)}. Therefore, in
the loop expansion of the kernel,
\begin{equation}
   \mathcal{K}_\Lambda(p)_{\mu\nu}^{ab}
   =\mathcal{K}_\Lambda^{(0)}(p)_{\mu\nu}^{ab}
   +g_\Lambda^2\mathcal{K}_\Lambda^{(1)}(p)_{\mu\nu}^{ab}+\dotsb,
\label{eq:(2.19)}
\end{equation}
we have, at tree level,
\begin{align}
   \mathcal{K}_\Lambda^{(0)}(p)_{\mu\nu}^{ab}
   =\delta^{ab}\left[
   (\delta_{\mu\nu}p^2-p_\mu p_\nu)\frac{1}{p^2/\Lambda^2+e^{-2tp^2}}
   +p_\mu p_\nu\frac{1}{p^2/\Lambda^2+\xi_\Lambda e^{-2\alpha_0tp^2}}\right],
\label{eq:(2.20)}
\end{align}
and, in the one-loop order,
\begin{align}
   &\mathcal{K}_\Lambda^{(1)}(p)_{\mu\nu}^{ab}
\notag\\
   &=-\frac{\Lambda^4}{\Lambda^{2\epsilon}g_\Lambda^4}
   G_{2t}^{\text{1-loop}}(p)_{\mu\nu}^{ab}
\notag\\
   &\qquad{}
   +\delta^{ab}\left[
   \frac{b_0}{\epsilon}
   (\delta_{\mu\nu}p^2-p_\mu p_\nu)\frac{e^{-2tp^2}}{(p^2/\Lambda^2+e^{-2tp^2})^2}
   +\frac{b_0-c_0}{\epsilon}
   p_\mu p_\nu\frac{\xi_\Lambda e^{-2\alpha_0tp^2}}
   {(p^2/\Lambda^2+\xi_\Lambda e^{-2\alpha_0tp^2})^2}\right],
\label{eq:(2.21)}
\end{align}
where the last term provides the one-loop level counter term.

To extract the coefficients $b_0$ and~$c_0$ from the two-point function
in~Eq.~\eqref{eq:(2.21)}, we now introduce a dimensionless function
$\Bar{\mathcal{K}}(\xi_\Lambda,p/\Lambda)_{\mu\nu}^{ab}$ by
\begin{align}
   &\Lambda^2\Bar{\mathcal{K}}(\xi_\Lambda,p/\Lambda)_{\mu\nu}^{ab}
\notag\\
   &:=\mathcal{K}_\Lambda^{(1)}(p)_{\mu\rho}^{ab}
   \left[
   \left(\delta_{\rho\nu}-\frac{p_\rho p_\nu}{p^2}\right)
   \frac{(p^2/\Lambda^2+e^{-2tp^2})^2}{e^{-2tp^2}}
   +\frac{p_\rho p_\nu}{p^2}
   \frac{(p^2/\Lambda^2+\xi_\Lambda e^{-2\alpha_0tp^2})^2}
   {\xi_\Lambda e^{-2\alpha_0tp^2}}
   \right]
\notag\\
   &=-\frac{\Lambda^4}{\Lambda^{2\epsilon}g_\Lambda^4}
   G_{2t}^{\text{1-loop}}(p)_{\mu\rho}^{ab}
   \left[
   \left(\delta_{\rho\nu}-\frac{p_\rho p_\nu}{p^2}\right)
   \frac{(p^2/\Lambda^2+e^{-2tp^2})^2}{e^{-2tp^2}}
   +\frac{p_\rho p_\nu}{p^2}
   \frac{(p^2/\Lambda^2+\xi_\Lambda e^{-2\alpha_0tp^2})^2}
   {\xi_\Lambda e^{-2\alpha_0tp^2}}
   \right]
\notag\\
   &\qquad{}
   +\delta^{ab}\left[
   \frac{b_0}{\epsilon}
   (\delta_{\mu\nu}p^2-p_\mu p_\nu)
   +\frac{b_0-c_0}{\epsilon}
   p_\mu p_\nu\right]
\notag\\
   &=:\Lambda^{2\epsilon}\int_\ell\,
   \mathcal{I}(\ell,p)_{\mu\nu}^{ab}
   +\delta^{ab}\left[
   \frac{b_0}{\epsilon}
   (\delta_{\mu\nu}p^2-p_\mu p_\nu)
   +\frac{b_0-c_0}{\epsilon}
   p_\mu p_\nu\right],
\label{eq:(2.22)}
\end{align}
where $\mathcal{I}(\ell,p)_{\mu\nu}^{ab}$ is defined as the integrand of the
one-loop momentum integral in the combination
$-\frac{\Lambda^4}{\Lambda^{4\epsilon}g_\Lambda^4}
   G_{2t}^{\text{1-loop}}(p)_{\mu\rho}^{ab}
   \left[
   \left(\delta_{\rho\nu}-\frac{p_\rho p_\nu}{p^2}\right)
   \frac{(p^2/\Lambda^2+e^{-2tp^2})^2}{e^{-2tp^2}}
   +\frac{p_\rho p_\nu}{p^2}
   \frac{(p^2/\Lambda^2+\xi_\Lambda e^{-2\alpha_0tp^2})^2}
   {\xi_\Lambda e^{-2\alpha_0tp^2}}
   \right]$.\footnote{%
The Feynman rule for a loop diagram provides an integrand of the loop
integration as a product of the propagators and vertices. For a set of
diagrams with an identical number of loops, we may define the integrand by the
sum of contributions from each diagrams. What we indicate
by~$\mathcal{I}(\ell,p)_{\mu\nu}^{ab}$ in~Eq.~\eqref{eq:(2.22)} is such an
integrand directly given by the Feynman rule
for~$G_{2t}^{\text{1-loop}}(p)_{\mu\rho}^{ab}$. See the last line
of~Eq.~\eqref{eq:(2.31)} as an example.} Taking the $\Lambda$~derivative of
both sides, we have
\begin{align}
   &-\Lambda\frac{\partial}{\partial\Lambda}
   \left[
   \Lambda^2\Bar{\mathcal{K}}(\xi_\Lambda,p/\Lambda)_{\mu\nu}^{ab}
   \right]
\notag\\
   &=-2\epsilon
   \left\{
   \Lambda^2\Bar{\mathcal{K}}(\xi_\Lambda,p/\Lambda)_{\mu\nu}^{ab}
   -\delta^{ab}\left[
   \frac{b_0}{\epsilon}
   (\delta_{\mu\nu}p^2-p_\mu p_\nu)
   +\frac{b_0-c_0}{\epsilon}
   p_\mu p_\nu\right]
   \right\}
\notag\\
   &\qquad{}
   +\Lambda^{2\epsilon}\int_\ell\,
   \left(-\Lambda\frac{\partial}{\partial\Lambda}\right)
   \mathcal{I}(\ell,p)_{\mu\nu}^{ab}.
\label{eq:(2.23)}
\end{align}
The function~$\mathcal{I}(\ell,p)_{\mu\nu}^{ab}$ depends on $\Lambda$, $p$,
and $\ell$, all are of mass dimension~$1$. Since the mass dimension
of~$\mathcal{I}(\ell,p)_{\mu\nu}^{ab}$ is~$-2$, from dimensional grounds, we have
\begin{equation}
   \left(
   \Lambda\frac{\partial}{\partial\Lambda}
   +p\cdot\frac{\partial}{\partial p}
   +\ell\cdot\frac{\partial}{\partial\ell}
   +2
   \right)
   \mathcal{I}(\ell,p)_{\mu\nu}^{ab}
   =0.
\label{eq:(2.24)}
\end{equation}
Using this in~Eq.~\eqref{eq:(2.23)},
\begin{align}
   &-\Lambda\frac{\partial}{\partial\Lambda}
   \left[
   \Lambda^2\Bar{\mathcal{K}}(\xi_\Lambda,p/\Lambda)_{\mu\nu}^{ab}
   \right]
\notag\\
   &=-2\epsilon
   \left\{
   \Lambda^2\Bar{\mathcal{K}}_\Lambda(\xi_\Lambda,p/\Lambda)_{\mu\nu}^{ab}
   -\delta^{ab}\left[
   \frac{b_0}{\epsilon}
   (\delta_{\mu\nu}p^2-p_\mu p_\nu)
   +\frac{b_0-c_0}{\epsilon}
   p_\mu p_\nu\right]
   \right\}
\notag\\
   &\qquad{}
   +\Lambda^{2\epsilon}\int_\ell\,
   \left(
   p\cdot\frac{\partial}{\partial p}
   +\ell\cdot\frac{\partial}{\partial\ell}+2
   \right)\mathcal{I}(\ell,p)_{\mu\nu}^{ab}.
\label{eq:(2.25)}
\end{align}
Then, taking terms quadratic in~$p$ from both sides, we have
\begin{align}
   0&=-2\epsilon
   \left\{
   \left.\Lambda^2\Bar{\mathcal{K}}_\Lambda(\xi_\Lambda,p/\Lambda)_{\mu\nu}^{ab}
   \right|_{O(p^2)}
   -\delta^{ab}\left[
   \frac{b_0}{\epsilon}
   (\delta_{\mu\nu}p^2-p_\mu p_\nu)
   +\frac{b_0-c_0}{\epsilon}
   p_\mu p_\nu\right]
   \right\}
\notag\\
   &\qquad{}
   +\Lambda^{2\epsilon}\int_\ell\,
   \left(
   \ell\cdot\frac{\partial}{\partial\ell}+4
   \right)
   \left.\mathcal{I}(\ell,p)_{\mu\nu}^{ab}\right|_{O(p^2)}.
\label{eq:(2.26)}
\end{align}
Although there remains the $\Lambda$~dependence through~$\xi_\Lambda$ on the
left-hand side, it is of higher order and can be neglected
in~Eq.~\eqref{eq:(2.26)}. The renormalization constants
in~Eq.~\eqref{eq:(2.13)} are determined so that
$\Bar{\mathcal{K}}_\Lambda(\xi_\Lambda,p/\Lambda)_{\mu\rho}^{ab}$ is finite
as~$\epsilon\to0$. Setting~$\epsilon\to0$ in~Eq.~\eqref{eq:(2.26)}, i.e,
$D\to4$, we thus finally have a useful representation,
\begin{align}
   \delta^{ab}\left[
   b_0(\delta_{\mu\nu}p^2-p_\mu p_\nu)
   +(b_0-c_0)p_\mu p_\nu\right]
   &=-\frac{1}{2}\int_\ell\,
   \left(
   \ell\cdot\frac{\partial}{\partial\ell}+4
   \right)
   \left.\mathcal{I}(\ell,p)_{\mu\nu}^{ab}
   \right|_{O(p^2)}
\notag\\
   &=-\frac{1}{2}\int_\ell\,\frac{\partial}{\partial\ell_\mu}
   \left[\ell_\mu
   \left.\mathcal{I}(\ell,p)_{\mu\nu}^{ab}
   \right|_{O(p^2)}\right]
\notag\\
   &=-\frac{1}{16\pi^2}\lim_{|\ell|\to\infty}(\ell^2)^2
   \left.\mathcal{I}(\ell,p)_{\mu\nu}^{ab}\right|_{O(p^2)}.
\label{eq:(2.27)}
\end{align}
Summarizing the above procedure, writing
\begin{equation}
   \left\langle A_\mu^{\prime a}(t,p)A_\nu^{\prime b}(t,q)\right\rangle^{\text{1-loop}}
   =(2\pi)^D\delta(p+q)G_{2t}^{\text{1-loop}}(p)_{\mu\nu}^{ab},
\label{eq:(2.28)}
\end{equation}
and defining the integrand~$\mathcal{I}(\ell,p)_{\mu\nu}^{ab}$ by
\begin{align}
   &-\frac{1}{\Lambda^{2\epsilon}g_\Lambda^4}
   G_{2t}^{\text{1-loop}}(p)_{\mu\rho}^{ab}
   \left[
   \left(\delta_{\rho\nu}-\frac{p_\rho p_\nu}{p^2}\right)
   \frac{(p^2+\Lambda^2e^{-2tp^2})^2}{e^{-2tp^2}}
   +\frac{p_\rho p_\nu}{p^2}
   \frac{(p^2+\xi_\Lambda\Lambda^2e^{-2\alpha_0tp^2})^2}
   {\xi_\Lambda e^{-2\alpha_0tp^2}}
   \right]
\notag\\
   &=
   \Lambda^{2\epsilon}\int_\ell\,\mathcal{I}(\ell,p)_{\mu\nu}^{ab},
\label{eq:(2.29)}
\end{align}
then the renormalization group coefficients $b_0$ and~$c_0$ can be read off from
the asymptotic behavior of $\mathcal{I}(\ell,p)_{\mu\nu}^{ab}|_{O(p^2)}$
at~$|\ell|\to\infty$, by~Eq.~\eqref{eq:(2.27)}.\footnote{%
Our present method is analogous to that of~Sect.~5.7
of~Ref.~\cite{Miyakawa:2021wus}.}

Our Feynman rule for~Eq.~\eqref{eq:(2.2)} is somewhat unconventional, because
of the Gaussian factors and flow or diffusion in~Eq.~\eqref{eq:(2.2)}. The
Feynman rule is summarized in~Appendix~\ref{sec:A}. According to that,
the one-loop two-point function~$G_{2t}^{\text{1-loop}}(p)_{\mu\nu}^{ab}$
in~Eq.~\eqref{eq:(2.28)} is given by the sum of diagrams depicted
in~Figs.~\ref{fig:1} and~\ref{fig:2}.\footnote{%
In drawing these diagrams, we have benefited from
Ti\textit{k}Z-Feynman~\cite{Ellis:2016jkw} and
Ti\textit{k}Z-Unhand~\cite{Dohse:2018vqo}.}
\begin{figure}[htbp]
\centering
\begin{subfigure}{0.3\columnwidth}
\centering
\begin{tikzpicture}[baseline=(a.base)]
\begin{feynhand}
\vertex [dot] (x) at (0,0) {};
\vertex [dot] (y) at (1.5,0) {};
\vertex [squaredot] (a) at (-1.5,0) {};
\vertex [squaredot] (b) at (3,0) {};
\propag [glu] (y) to [out=100, in=80, looseness=1.5] (x);
\propag [glu] (y) to [out=-100, in=-80, looseness=1.5] (x);
\propag [glu] (x) to (a);
\propag [glu] (b) to (y);
\end{feynhand}
\end{tikzpicture}
\caption{}
\label{fig:}
\end{subfigure}
\hspace{0mm}
\begin{subfigure}{0.3\columnwidth}
\centering
\begin{tikzpicture}[baseline=(a.base)]
\begin{feynhand}
\vertex [dot] (x) at (0,0) {};
\vertex [squaredot] (a) at (-1.5,0) {};
\vertex [squaredot] (b) at (1.5,0) {};
\propag [glu] (x) to [in=130, out=50, looseness=60] (x);
\propag [glu] (x) to (a);
\propag [glu] (b) to (x);
\end{feynhand}
\end{tikzpicture}
\caption{}
\label{fig:}
\end{subfigure}
\hspace{0mm}
\begin{subfigure}{0.3\columnwidth}
\centering
\begin{tikzpicture}[baseline=(a.base)]
\begin{feynhand}
\vertex [dot] (x) at (0,0) {};
\vertex [dot] (y) at (1.5,0) {};
\vertex [squaredot] (a) at (-1.5,0) {};
\vertex [squaredot] (b) at (3,0) {};
\propag [gho] (y) to [out=100, in=80, looseness=1.5] (x);
\propag [gho] (y) to [out=-100, in=-80, looseness=1.5] (x);
\propag [glu] (x) to (a);
\propag [glu] (b) to (y);
\end{feynhand}
\end{tikzpicture}
\caption{}
\label{fig:}
\end{subfigure}
\\[8mm]
\centering
\begin{subfigure}{0.3\columnwidth}
\centering
\begin{tikzpicture}[baseline=(a.base)]
\begin{feynhand}
\vertex [ringblob,minimum height=3.5mm,minimum width=3.5mm] (x) at (0,0) {};
\vertex [dot] (y) at (1.5,0) {};
\vertex [squaredot] (a) at (-1.5,0) {};
\vertex [squaredot] (b) at (3,0) {};
\propag [glu] (y) to [out=100, in=80, looseness=1.5] (x);
\propag [glu] (y) to [out=-100, in=-80, looseness=1.5] (x);
\propag [fer] (x) to (a);
\propag [glu] (b) to (y);
\end{feynhand}
\end{tikzpicture}
\caption{}
\label{fig:}
\end{subfigure}
\hspace{0mm}
\begin{subfigure}{0.3\columnwidth}
\centering
\begin{tikzpicture}[baseline=(a.base)]
\begin{feynhand}
\vertex [ringblob,minimum height=3.5mm,minimum width=3.5mm] (x) at (0,0) {};
\vertex [ringblob,minimum height=3.5mm,minimum width=3.5mm] (y) at (1.5,0) {};
\vertex [squaredot] (a) at (-1.5,0) {};
\vertex [squaredot] (b) at (3,0) {};
\propag [glu] (y) to [out=100, in=80, looseness=1.5] (x);
\propag [glu] (y) to [out=-100, in=-80, looseness=1.5] (x);
\propag [fer] (x) to (a);
\propag [fer] (y) to (b);
\end{feynhand}
\end{tikzpicture}
\caption{}
\label{fig:}
\end{subfigure}
\hspace{0mm}
\begin{subfigure}{0.3\columnwidth}
\centering
\begin{tikzpicture}[baseline=(a.base)]
\begin{feynhand}
\vertex [ringblob,minimum height=3.5mm,minimum width=3.5mm] (x) at (0,0) {};
\vertex [squaredot] (a) at (-1.5,0) {};
\vertex [squaredot] (b) at (1.5,0) {};
\propag [glu] (x) to [in=130, out=50, looseness=25] (x);
\propag [fer] (x) to (a);
\propag [glu] (b) to (x);
\end{feynhand}
\end{tikzpicture}
\caption{}
\label{fig:}
\end{subfigure}
\\[8mm]
\centering
\begin{subfigure}{0.3\columnwidth}
\centering
\begin{tikzpicture}[baseline=(a.base)]
\begin{feynhand}
\vertex [ringblob,minimum height=3.5mm,minimum width=3.5mm] (x) at (0,0) {};
\vertex [ringblob,minimum height=3.5mm,minimum width=3.5mm] (y) at (1.5,0) {};
\vertex [squaredot] (a) at (-1.5,0) {};
\vertex [squaredot] (b) at (3,0) {};
\propag [glu] (y) to [out=100, in=80, looseness=1.5] (x);
\propag [fer] (y) to (x);
\propag [fer] (x) to (a);
\propag [glu] (y) to (b);
\end{feynhand}
\end{tikzpicture}
\caption{}
\label{fig:}
\end{subfigure}
\caption{One-loop level diagrams which contribute to the two-point
function~\eqref{eq:(2.28)}.}
\label{fig:1}
\end{figure}
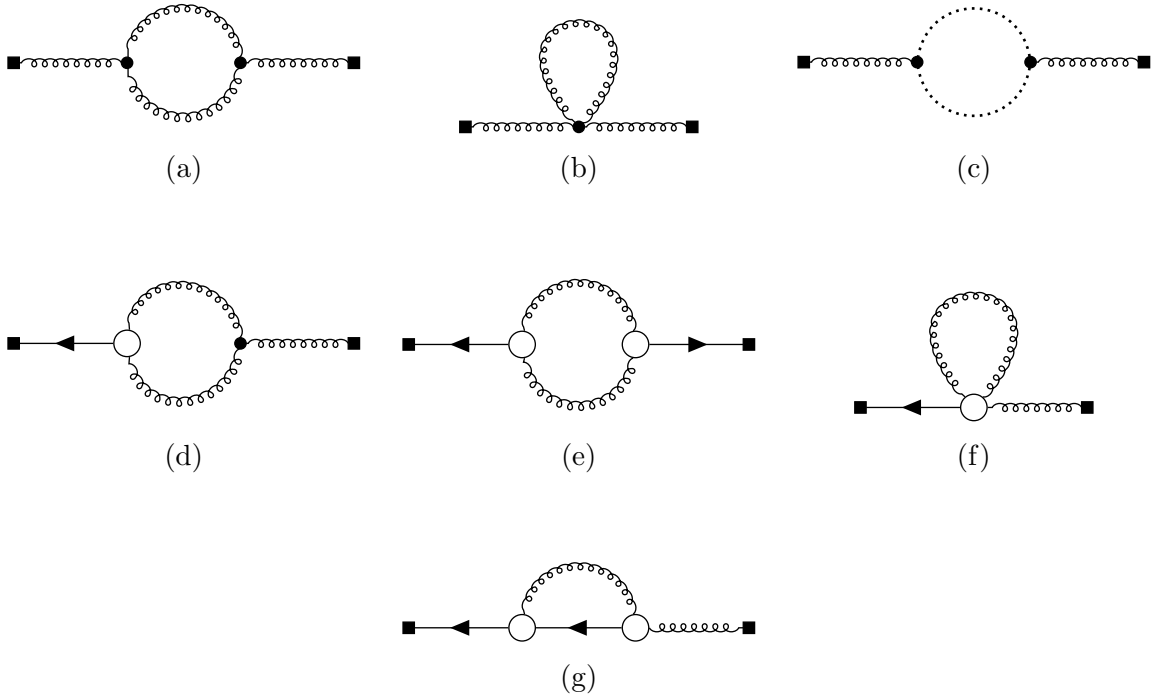
\begin{figure}[htbp]
\centering
\begin{subfigure}{0.3\columnwidth}
\centering
\begin{tikzpicture}[baseline=(a.base)]
\begin{feynhand}
\vertex [grayblob,minimum height=3.5mm,minimum width=3.5mm] (x) at (0,0) {};
\vertex [dot] (y) at (1.5,0) {};
\vertex [squaredot] (a) at (-1.5,0) {};
\vertex [squaredot] (b) at (3,0) {};
\propag [glu] (y) to [out=100, in=80, looseness=1.5] (x);
\propag [glu] (y) to [out=-100, in=-80, looseness=1.5] (x);
\propag [glu] (x) to (a);
\propag [glu] (b) to (y);
\end{feynhand}
\end{tikzpicture}
\caption{}
\label{fig:}
\end{subfigure}
\hspace{0mm}
\begin{subfigure}{0.3\columnwidth}
\centering
\begin{tikzpicture}[baseline=(a.base)]
\begin{feynhand}
\vertex [grayblob,minimum height=3.5mm,minimum width=3.5mm] (x) at (0,0) {};
\vertex [grayblob,minimum height=3.5mm,minimum width=3.5mm] (y) at (1.5,0) {};
\vertex [squaredot] (a) at (-1.5,0) {};
\vertex [squaredot] (b) at (3,0) {};
\propag [glu] (y) to [out=100, in=80, looseness=1.5] (x);
\propag [glu] (y) to [out=-100, in=-80, looseness=1.5] (x);
\propag [glu] (x) to (a);
\propag [glu] (b) to (y);
\end{feynhand}
\end{tikzpicture}
\caption{}
\label{fig:}
\end{subfigure}
\hspace{0mm}
\begin{subfigure}{0.3\columnwidth}
\centering
\begin{tikzpicture}[baseline=(a.base)]
\begin{feynhand}
\vertex [grayblob,minimum height=3.5mm,minimum width=3.5mm] (x) at (0,0) {};
\vertex [dot] (y) at (1.5,0) {};
\vertex [squaredot] (a) at (-1.5,0) {};
\vertex [squaredot] (b) at (3,0) {};
\propag [gho] (y) to [out=100, in=80, looseness=1.5] (x);
\propag [gho] (y) to [out=-100, in=-80, looseness=1.5] (x);
\propag [glu] (x) to (a);
\propag [glu] (b) to (y);
\end{feynhand}
\end{tikzpicture}
\caption{}
\label{fig:}
\end{subfigure}
\\[8mm]
\centering
\begin{subfigure}{0.3\columnwidth}
\centering
\begin{tikzpicture}[baseline=(a.base)]
\begin{feynhand}
\vertex [grayblob,minimum height=3.5mm,minimum width=3.5mm] (x) at (0,0) {};
\vertex [grayblob,minimum height=3.5mm,minimum width=3.5mm] (y) at (1.5,0) {};
\vertex [squaredot] (a) at (-1.5,0) {};
\vertex [squaredot] (b) at (3,0) {};
\propag [gho] (y) to [out=100, in=80, looseness=1.5] (x);
\propag [gho] (y) to [out=-100, in=-80, looseness=1.5] (x);
\propag [glu] (x) to (a);
\propag [glu] (b) to (y);
\end{feynhand}
\end{tikzpicture}
\caption{}
\label{fig:}
\end{subfigure}
\hspace{0mm}
\begin{subfigure}{0.3\columnwidth}
\centering
\begin{tikzpicture}[baseline=(a.base)]
\begin{feynhand}
\vertex [ringblob,minimum height=3.5mm,minimum width=3.5mm] (x) at (0,0) {};
\vertex [grayblob,minimum height=3.5mm,minimum width=3.5mm] (y) at (1.5,0) {};
\vertex [squaredot] (a) at (-1.5,0) {};
\vertex [squaredot] (b) at (3,0) {};
\propag [glu] (y) to [out=100, in=80, looseness=1.5] (x);
\propag [glu] (y) to [out=-100, in=-80, looseness=1.5] (x);
\propag [fer] (x) to (a);
\propag [glu] (b) to (y);
\end{feynhand}
\end{tikzpicture}
\caption{}
\label{fig:}
\end{subfigure}
\hspace{0mm}
\begin{subfigure}{0.3\columnwidth}
\centering
\begin{tikzpicture}[baseline=(a.base)]
\begin{feynhand}
\vertex [grayblob,minimum height=3.5mm,minimum width=3.5mm] (x) at (0,0) {};
\vertex [squaredot] (a) at (-1.5,0) {};
\vertex [squaredot] (b) at (1.5,0) {};
\propag [gho] (x) to [in=130, out=50, looseness=25] (x);
\propag [glu] (x) to (a);
\propag [glu] (b) to (x);
\end{feynhand}
\end{tikzpicture}
\caption{}
\label{fig:}
\end{subfigure}
\caption{One-loop level diagrams which contribute to the two-point
function~\eqref{eq:(2.28)}. Shaded blob vertices are the effect of the
mass term arising from the Gaussian factors.}
\label{fig:2}
\end{figure}
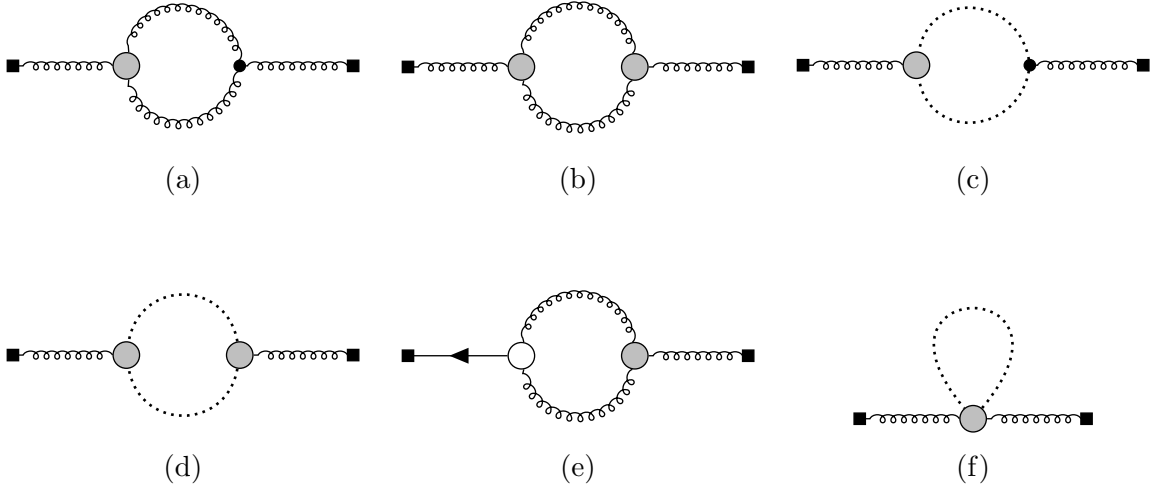
Recall that the mass dimension of the
function~$\mathcal{I}(\ell,p)_{\mu\nu}^{ab}|_{O(p^2)}$ in~Eq.~\eqref{eq:(2.27)}
is~$-2$. For the limit~$\lim_{|\ell|\to\infty}(\ell^2)^2%
\left.\mathcal{I}(\ell,p)_{\mu\nu}^{ab}\right|_{O(p^2)}$ in~Eq.~\eqref{eq:(2.27)} to
be non-zero, therefore, $\left.\mathcal{I}(\ell,p)_{\mu\nu}^{ab}\right|_{O(p^2)}$
should behave as~$\delta_{\mu\nu}p^2/(\ell^2)^2$ or~$p_\mu p_\nu/(\ell^2)^2$ with
\emph{dimensionless\/} coefficients. If the interaction vertex is proportional
to~$\Lambda^2$, as the shaded blob vertices in~Fig.~\ref{fig:2}, given
by~Eqs.~\eqref{eq:(A29)}--\eqref{eq:(A32)}, therefore, the diagram does not
contribute to the limit~\eqref{eq:(2.27)}. This shows that the coefficients
$b_0$ and~$c_0$ can be determined without computing diagrams
in~Fig.~\ref{fig:2}. Of course, to determine the whole Wilson action in the
one-loop level, one also has to compute diagrams in~Fig.~\ref{fig:2}. All the
diagrams in~Fig.~\ref{fig:1} potentially contribute to the
limit~\eqref{eq:(2.27)}.

Let us illustrate the calculation of~$G_{2t}^{\text{1-loop}}(p)_{\mu\nu}^{ab}$
in~Eq.~\eqref{eq:(2.28)}, taking diagram~(d) of~Fig.~\ref{fig:1} as an
example. The calculation of other diagrams is reported in~Appendix~\ref{sec:B}.
Following the Feynman rule in~Appendix~\ref{sec:A}, we
have\footnote{%
In one-loop calculations, we set $g_0\to g_\Lambda$ and~$\xi_0\to\xi_\Lambda$
because the differences are of higher orders.}
\begin{align}
   &G_{2t}^{\text{1-loop}}(p)_{\mu\nu}^{ab}[\text{diagram~(d) of~Fig.~\ref{fig:1}}]
\notag\\
   &=ig_\Lambda^4\int_0^tds\,K_{t-s}(p)_{\mu\rho}\int_\ell\,
   X^{(2,0)}(p,\ell,-\ell-p)_{\rho\alpha\beta}^{acd}
   D_s(\ell)_{\alpha\gamma}D_s(\ell+p)_{\beta\delta}
\notag\\
   &\qquad{}
   \times
   f^{bcd}
   \left[
   (2\ell+p)_\sigma\delta_{\gamma\delta}
   -(\ell+2p)_\gamma\delta_{\delta\sigma}
   -(\ell-p)_\delta\delta_{\sigma\gamma}
   \right]
   D_t(p)_{\sigma\nu}.
\label{eq:(2.30)}
\end{align}
After substituting this into~Eq.~\eqref{eq:(2.29)}, we want to find the term
quadratic in~$p$ for~Eq.~\eqref{eq:(2.27)}. For this, we can set $p\to0$
in~Eq.~\eqref{eq:(2.30)} as (noting that
$\lim_{p\to0}K_{t-s}(p)_{\mu\rho}=\delta_{\mu\rho}$)
\begin{align}
   &G_{2t}(p)_{\mu\nu}^{ab}[\text{diagram~(d) of~Fig.~\ref{fig:1}}]
\notag\\
   &\to
   ig_\Lambda^4\int_0^tds\,\int_\ell\,
   X^{(2,0)}(0,\ell,-\ell)_{\mu\alpha\beta}^{acd}
   D_s(\ell)_{\alpha\gamma}D_s(\ell)_{\beta\delta}
   f^{bcd}
   \left(
   2\ell_\sigma\delta_{\gamma\delta}
   -\ell_\gamma\delta_{\delta\sigma}
   -\ell_\delta\delta_{\sigma\gamma}
   \right)
   D_t(p)_{\sigma\nu}
\notag\\
   &\to
   g_\Lambda^4C_A\delta^{ab}\int_0^tds\,\int_\ell\,
   \left[
   4(D-1)\frac{\ell_\mu\ell_\sigma}{\ell^2}\frac{e^{-2s\ell^2}}{\ell^2}
   +2(\alpha_0+1)\xi_\Lambda
   \left(\delta_{\mu\sigma}-\frac{\ell_\mu\ell_\sigma}{\ell^2}\right)
   \frac{e^{-(\alpha_0+1)s\ell^2}}{\ell^2}
   \right]
\notag\\
   &\qquad{}
   \times D_t(p)_{\sigma\nu}
\notag\\
   &\to
   g_\Lambda^4C_A\delta^{ab}\int_\ell\,
   \left[
   2(D-1)\frac{\ell_\mu\ell_\sigma}{\ell^2}\frac{1}{(\ell^2)^2}
   +2\xi_\Lambda\left(\delta_{\mu\sigma}-\frac{\ell_\mu\ell_\sigma}{\ell^2}\right)
   \frac{1}{(\ell^2)^2}
   \right]
   D_t(p)_{\sigma\nu}
\notag\\
   &=g_\Lambda^4C_A\delta^{ab}\times 2(1+\xi_\Lambda)\left(1-\frac{1}{D}\right)
   \int_\ell\,\frac{1}{(\ell^2)^2}D_t(p)_{\mu\nu},
\label{eq:(2.31)}
\end{align}
where we have kept only the terms that can contribute in the
limit~$|\ell|\to\infty$ in~Eq.~\eqref{eq:(2.27)}. From~Eq.~\eqref{eq:(2.29)},
this contribute to $\mathcal{I}(\ell,p)_{\mu\nu}^{ab}|_{O(p^2)}$ by for~$D\to4$,
\begin{equation}
   \mathcal{I}(\ell,p)_{\mu\nu}^{ab}[\text{diagram~(d) of~Fig.~\ref{fig:1}}]
   |_{O(p^2)}
   \to-C_A\delta^{ab}\frac{3}{2}(1+\xi_\Lambda)
   \frac{1}{(\ell^2)^2}\delta_{\mu\nu}p^2.
\label{eq:(2.32)}
\end{equation}
Thus, from Eq.~\eqref{eq:(2.27)}, this diagram contributes
to~$b_0=C_A/(16\pi^2)(3/2)(1+\xi_\Lambda)$ and~$c_0=0$.

Other diagrams in~Fig.~\ref{fig:1} can be evaluated in a similar way;
see~Appendix~\ref{sec:B}. In~Table~\ref{table:1}, we summarize the contribution
of each diagram.
\begin{table}[htbp]
\caption{Contribution of each diagram in~Fig.~\ref{fig:1} to $b_0$ and~$c_0$
in~Eq.~\eqref{eq:(2.14)} in unit of~$C_A/(16\pi^2)$.}
\label{table:1}
\begin{center}
\begin{tabular}{ccc}
\toprule
diagram&$b_0$&$c_0$\\
\midrule
(a)&$\frac{25}{12}-\frac{1}{2}\xi_\Lambda$&$\frac{25}{12}-\frac{1}{4}\xi_\Lambda$\\
(b)&$0$&$0$\\
(c)&$\frac{1}{12}$&$\frac{1}{12}-\frac{1}{4}\xi_\Lambda$\\
(d)&$\frac{3}{2}+\frac{3}{2}\xi_\Lambda$&$0$\\
(e)&$0$&$0$\\
(f)&$-\frac{9}{4}-\frac{3}{4}\frac{\xi_\Lambda}{\alpha_0}$&$0$\\
(g)&$\frac{9}{4}-\xi_\Lambda+\frac{3}{4}\frac{\xi_\Lambda}{\alpha_0}$&$0$\\
\bottomrule
\end{tabular}
\end{center}
\end{table}
Gathering all contributions in~Table~\ref{table:1}, we finally obtain
\begin{equation}
   b_0=\frac{1}{16\pi^2}C_A\frac{11}{3},
   \qquad c_0=\frac{1}{16\pi^2}C_A
   \left(\frac{13}{6}-\frac{1}{2}\xi_\Lambda\right).
\label{eq:(2.33)}
\end{equation}
This coincides with the well-known correct result. By Eqs.~\eqref{eq:(2.13)}
and~\eqref{eq:(2.14)}, these tell the running under the change of~$\Lambda$,
\begin{equation}
   -\Lambda\frac{d}{d\Lambda}g_\Lambda
   =\epsilon g_\Lambda+b_0 g_\Lambda^3+O(g_\Lambda^5),\qquad
   -\Lambda\frac{d}{d\Lambda}\xi_\Lambda
   =\left[-2c_0g_\Lambda^2+O(g_\Lambda^4)\right]\xi_\Lambda.
\label{eq:(2.34)}
\end{equation}

We note that the numbers in~Table~\ref{table:1} had already been obtained
in~Sect.~5.1 of~Ref.~\cite{Luscher:2011bx} as one-loop divergent parts in the
self-energy of the flowed gauge field in the gradient flow formalism with the
dimensional regularization. The diagrams in~Fig.~\ref{fig:1} are completely
identical to diagrams in~Fig.~4 of~Ref.~\cite{Luscher:2011bx}. Actually, what
we have evaluated in~Eq.~\eqref{eq:(2.27)} is the logarithmic divergent part
in the two-point function
$\langle A_\mu^{\prime a}(t,x)A_\nu^{\prime b}(t,y)\rangle_S$. We expect these
coincidences from general grounds because, as discussed
in~Ref.~\cite{Sonoda:2025tyu}, the Wilson action~$S_\Lambda$ in GFERG reproduces
correlation functions of the flowed fields, $A_\mu^{\prime a}(t,x)$
and~$c^{\prime a}(t,x)$, in the gradient flow formalism up to contact terms under
the identification~\eqref{eq:(2.8)}. As discussed
in~Ref.~\cite{Sonoda:2025tyu}, Eqs.~\eqref{eq:(2.2)} and~\eqref{eq:(2.8)} imply
the equality
\begin{align}
   &\biggl\langle
   \exp\left[
   -\frac{1}{2\Lambda^2}\int d^Dx\,
   \frac{\delta^2}{\delta A_\mu^a(x)\delta A_\mu^a(x)}\right]
   \exp\left[
   \frac{1}{\Lambda^2}\int d^Dx\,
   \frac{\delta}{\delta c^a(x)}\frac{\delta}{\delta\Bar{c}^a(x)}\right]
\notag\\
   &\qquad{}
   \times
   A_{\mu_1}^{a_1}(x_1)\dotsb A_{\mu_N}^{a_N}(x_N)
   c^{b_1}(y_1)\dotsb c^{b_M}(y_M)
   \Bar{c}^{c_1}(z_1)\dotsb\Bar{c}^{c_L}(z_L)
   \biggr\rangle_{S_\Lambda}
\notag\\
   &=Z_{\Bar{c}}(\Lambda)^L
   (\Lambda^\epsilon g_\Lambda)^{-N-M-L}
\notag\\
   &\qquad{}
   \times\left\langle
   A_{\mu_1}^{a_1}(t,x_1)\dotsb A_{\mu_N}^{a_N}(t,x_N)
   c^{b_1}(t,y_1)\dotsb c^{b_M}(t,y_M)
   \Bar{c}^{c_1}(z_1)\dotsb\Bar{c}^{c_L}(z_L)
   \right\rangle_S,
\label{eq:(2.35)}
\end{align}
where the expectation value on the left-hand side is computed by employing the
Wilson action~$S_\Lambda$, whereas the expectation value on the right-hand side
is computed by the bare action~$S$. Then, an expansion of the functional
derivative operators in the expectation value on the left-hand side produces
terms such as
\begin{align}
   \delta^{a_1a_2}\delta_{\mu_1\mu_2}\delta(x_1-x_2)\biggl\langle
   A_{\mu_3}^{a_3}(x_3)\dotsb A_{\mu_N}^{a_N}(x_N)
   c^{b_1}(y_1)\dotsb c^{b_M}(y_M)
   \Bar{c}^{c_1}(z_1)\dotsb\Bar{c}^{c_L}(z_L)
   \biggr\rangle_{S_\Lambda}.
\label{eq:(2.36)}
\end{align}
These are the above-mentioned contact terms.\footnote{%
We are not aware of a general proof demonstrating that such contact terms in
coordinate space have no physical effect at all. However, at least, they do not
contribute to the S-matrix obtained in the long–distance limit of correlation
functions. On the other hand, in the ERG context, it is
known~\cite{Sonoda:2015bla} that correlation functions exhibiting covariant
behavior under the scale transformation generically contain such contact terms
compared with ordinary correlation functions. Therefore , the appearance of
contact terms in~Eq.~\eqref{eq:(2.35)} is entirely natural from the ERG
perspective.} Equation~\eqref{eq:(2.35)} shows that, therefore, the gauge
potential two-point functions in the GFERG framework and the gradient flow
formalism share the common UV divergent parts. More generally, since terms such
as~Eq.~\eqref{eq:(2.36)} are simply correlation functions computed
by~$S_\Lambda$, the conditions for the UV finiteness are common for the present
ERG framework and the gradient flow formalism.

\subsection{Higher orders}
\label{sec:2.3}
As mentioned just now, correlation functions computed by employing the Wilson
action~$S_\Lambda$ coincide with correlation functions of flowed fields in the
gradient flow formalism up to contact terms~\cite{Sonoda:2025tyu}. The
requirement of the finiteness of the Wilson action is thus equivalent to the
renormalization in the original Yang--Mills theory. In perturbation theory, we
may assume the dimensional regularization that regularizes the bare action~$S$
in a BRST invariant manner. Thus, we expect that the the renormalization group
functions in the Minimal Subtraction (MS) scheme, for instance, make the Wilson
action in GFERG finite in all orders of perturbation theory. This implies that
the GFERG reproduces the correct conventional renormalization group functions
in all orders in perturbation theory, up to the scheme ambiguity (any
renormalization scheme renders the Wilson action finite).

\section{Conclusion}
\label{sec:3}
In this paper, we have computed the one-loop renormalization group coefficients
in the pure Yang--Mills theory in GFERG. We obtained these by directly
expanding the integral representation of the Wilson action in terms of the
bare action~\eqref{eq:(2.2)} and imposing the finiteness of the Wilson action
in the $\Lambda_0\to\infty$ (or more precisely $\epsilon\to0$) limit. By
construction, the resulting Wilson action should fulfill the GFERG flow
equation~\eqref{eq:(2.9)} and the BRST WT identity~\eqref{eq:(2.11)}
simultaneously, although in the present paper, we have explicitly computed only
a small portion of the Wilson action, which provides the information of the
renormalization group coefficients. The present calculation reproducing the
correct one-loop renormalization group coefficients illustrates an internal
consistency of GFERG.

\section*{Acknowledgments}
We are grateful to Hidenori Sonoda for his collaboration in the early stage of
this work.
The work of H.S. was partially supported by Japan Society for the Promotion of
Science (JSPS) Grant-in-Aid for Scientific Research, JP23K03418.

\appendix
\section{Feynman rule for correlation functions under the functional
integral~\eqref{eq:(2.2)}}
\label{sec:A}
In this Appendix, we present the Feynman rule to compute correlation
functions under the functional integral~\eqref{eq:(2.2)}. The Feynman rule is
almost identical to that in the gradient flow
formalism~\cite{Luscher:2010iy,Luscher:2011bx}, except for the effect of the
Gaussian factors in~Eq.~\eqref{eq:(2.2)}. We therefore follow the notational
convention of~Refs.~\cite{Luscher:2010iy,Luscher:2011bx} as much as possible.

\subsection{Perturbative solution to flow equations}
\label{sec:A.1}
First, in~Eq.~\eqref{eq:(2.2)}, we need to represent the flowed or diffused
fields, $A_\mu'(t,x)$ and~$c'(t,x)$, in terms of the integration variables,
$A_\mu'(x)$ and~$c'(x)$. This is done by perturbatively solving the flow
equations~\eqref{eq:(2.5)} and~\eqref{eq:(2.7)}. Equation~\eqref{eq:(2.5)}
can be expressed as the integral equation~\cite{Luscher:2010iy,Luscher:2011bx},
\begin{equation}
   A_\mu^{\prime a}(t,x)
   =\int d^Dy\left[
   K_t(x-y)_{\mu\nu}A_\nu^{\prime a}(y)
   +\int_0^tds\,K_{t-s}(x-y)_{\mu\nu}R_\nu^{\prime a}(s,y)
   \right],
\label{eq:(A1)}
\end{equation}
where $K_t(x)$ is the hear kernel,
\begin{equation}
   K_t(x)_{\mu\nu}
   :=\int_p\,e^{ipx}
   \left[\left(\delta_{\mu\nu}-\frac{p_\mu p_\nu}{p^2}\right)e^{-tp^2}
   +\frac{p_\mu p_\nu}{p^2}e^{-\alpha_0tp^2}\right],
\label{eq:(A2)}
\end{equation}
and non-linear terms are given by
\begin{equation}
   R_\mu^{\prime a}
   =f^{abc}\left[
   2A_\nu^{\prime b}\partial_\nu A_\mu^{\prime c}
   -A_\nu^{\prime b}\partial_\mu A_\nu^{\prime c}
   +(\alpha_0-1)A_\mu^{\prime b}\partial_\nu A_\nu^{\prime c}
   +f^{cde}A_\nu^{\prime b}A_\nu^{\prime d}A_\mu^{\prime e}\right].
\label{eq:(A3)}
\end{equation}
We represent the non-linear terms as, in momentum space,
\begin{align}
   R_\mu^{\prime a}(t,p)
   &=\sum_{n=2}^3
   \frac{1}{n!}\int_{q_1}\dotsb\int_{q_n}(2\pi)^D\delta(p+q_1+\dotsb+q_n)
\notag\\
   &\qquad{}
   \times
   X^{(n,0)}(p,q_1,\dotsc, q_n)_{\mu\nu_1,\dotsb\nu_n}^{ab_1\dotsb b_n}
   A_{\nu_1}^{\prime b_1}(t,-q_1)\dotsb A_{\nu_n}^{\prime b_n}(t,-q_n).
\label{eq:(A4)}
\end{align}
These ``flow vertices'' $X^{(n,0)}$ ($n=2$, $3$), are diagrammatically
represented by an open circle, and are explicitly given
by~\cite{Luscher:2010iy,Luscher:2011bx},
\begin{align}
&\begin{tikzpicture}[baseline=(a.base)]
\begin{feynhand}
\vertex [ringblob,minimum height=3.5mm,minimum width=3.5mm] (a) at (0,0) {};
\vertex (b) at (-2,0) {$(p,\mu,a)$};
\vertex (c) at (1.4,1.4) {$(q,\nu,b)$};
\vertex (d) at (1.4,-1.4) {$(r,\rho,c)$};
\propag [fer, revmom'={$p$}] (a) to (b);
\propag [fer, mom'={$q$}] (c) to (a);
\propag [fer, mom={$r$}] (d) to (a);
\end{feynhand}
\end{tikzpicture}
   =\frac{1}{2!}X^{(2,0)}(p,q,r)_{\mu\nu\rho}^{abc}
\notag\\
   &=
   \frac{i}{2!}f^{abc}
   \left[(r-q)_\mu\delta_{\nu\rho}
   +2q_\rho\delta_{\mu\nu}-2r_\nu\delta_{\mu\rho}
   +(\alpha_0-1)(q_\nu\delta_{\mu\rho}-r_\rho\delta_{\mu\nu})\right],
\label{eq:(A5)}
\end{align}
and
\begin{align}
&\begin{tikzpicture}[baseline=(a.base)]
\begin{feynhand}
\vertex [ringblob,minimum height=3.5mm,minimum width=3.5mm] (a) at (0,0) {};
\vertex (b) at (-2,0) {$(p,\mu,a)$};
\vertex (c) at (1.4,1.4) {$(q,\nu,b)$};
\vertex (d) at (2,0) {$(r,\rho,c)$};
\vertex (e) at (1.4,-1.4) {$(s,\sigma,d)$};
\propag [fer, revmom'={$p$}] (a) to (b);
\propag [fer, mom'={$q$}] (c) to (a);
\propag [fer, mom'={$r$}] (d) to (a);
\propag [fer, mom={$s$}] (e) to (a);
\end{feynhand}
\end{tikzpicture}
   =\frac{1}{3!}X^{(3,0)}(p,q,r,s)_{\mu\nu\rho\sigma}^{abcd}
\notag\\
   &=\frac{1}{3!}\left\{
   f^{abe}f^{cde}
   \left[
   \delta_{\mu\sigma}\delta_{\nu\rho}-\delta_{\mu\rho}\delta_{\sigma\nu}\right]
   +f^{ade}f^{bce}
   \left[
   \delta_{\mu\rho}\delta_{\sigma\nu}-\delta_{\mu\nu}\delta_{\rho\sigma}\right]
   +f^{ace}f^{dbe}
   \left[
   \delta_{\mu\nu}\delta_{\rho\sigma}-\delta_{\mu\sigma}\delta_{\nu\rho}\right]
   \right\}.
\label{eq:(A6)}
\end{align}
In these ``flow line diagrams'', the arrow represents the direction of the
flow time, from the past to the future. 

Then, by iteratively solving the integral equation~\eqref{eq:(A1)}, we have a
perturbative expansion for the flowed field~$A_\mu'(t,x)$ in terms of the
initial value~$A_\mu'(x)$. The first term is given by
\begin{equation}
\begin{tikzpicture}
\begin{feynhand}
\vertex [squaredot] (a) at (0,0) {};
\vertex [crossdot] (b) at (2,0) {};
\propag [fer, revmom'={$p$}] (b) to (a);
\end{feynhand}
\end{tikzpicture}
   =K_t(p)_{\mu\nu}A_\nu^{\prime a}(p),
\label{eq:(A7)}
\end{equation}
where and in what follows the small square and the cross represent the flowed
field~$A_\mu^{\prime a}(t,p)$ and the initial
value~$A_\mu^{\prime a}(t=0,p)=A_\mu^{\prime a}(0,p)$, respectively.

Terms in the first order in~$R_\nu^{\prime \nu}(s,y)$ are given by
\begin{align}
&\begin{tikzpicture}[baseline=(a.base)]
\begin{feynhand}
\vertex [ringblob,minimum height=3.5mm,minimum width=3.5mm] (a) at (0,0) {};
\vertex [squaredot] (b) at (-1.5,0) {};
\vertex [crossdot] (c) at (1.1,1.1) {};
\vertex [crossdot] (d) at (1.1,-1.1) {};
\propag [fer, revmom'={$p$}] (a) to (b);
\propag [fer, mom'={$q$}] (c) to (a);
\propag [fer, mom={$-p-q$}] (d) to (a);
\end{feynhand}
\end{tikzpicture}
\notag\\
   &=\frac{1}{2!}\int_0^tds\,\int_qK_{t-s}(p)_{\mu\alpha}
   X^{(2,0)}(p,q,-p-q)_{\alpha\beta\gamma}^{abc}
   K_s(q)_{\beta\nu}A_\nu^{\prime b}(-q)K_s(p+q)_{\gamma\rho}A_\rho^{\prime c}(-p-q),
\label{eq:(A8)}
\end{align}
and
\begin{align}
&\begin{tikzpicture}[baseline=(a.base)]
\begin{feynhand}
\vertex [ringblob,minimum height=3.5mm,minimum width=3.5mm] (a) at (0,0) {};
\vertex [squaredot] (b) at (-1.5,0) {};
\vertex [crossdot] (c) at (1.1,1.1) {};
\vertex [crossdot] (d) at (1.5,0) {};
\vertex [crossdot] (e) at (1.1,-1.1) {};
\propag [fer, revmom'={$p$}] (a) to (b);
\propag [fer, mom'={$q$}] (c) to (a);
\propag [fer, mom'={$r$}] (d) to (a);
\propag [fer, mom={$-p-q-r$}] (e) to (a);
\end{feynhand}
\end{tikzpicture}
\notag\\
   &=\frac{1}{3!}\int_0^tds\,\int_{q,r}K_{t-s}(p)_{\mu\alpha}
   X^{(3,0)}(p,q,r,-p-q-r)_{\alpha\beta\gamma\delta}^{abcd}
\notag\\
   &\qquad{}
   \times K_s(q)_{\beta\nu}A_\nu^{\prime b}(-q)K_s(r)_{\gamma\rho}A_\rho^{\prime c}(-r)
   K_s(p+q+r)_{\delta\sigma}A_\rho^{\prime d}(p+q+r).
\label{eq:(A9)}
\end{align}

Then, the first iteration in~Eq.~\eqref{eq:(A1)} gives 
\begin{align}
&\begin{tikzpicture}[baseline=(b.base)]
\begin{feynhand}
\vertex [squaredot] (a) at (-1.5,0) {};
\vertex [ringblob,minimum height=3.5mm,minimum width=3.5mm] (b) at (0,0) {};
\vertex [ringblob,minimum height=3.5mm,minimum width=3.5mm] (c) at (1.5,0) {};
\vertex [crossdot] (d) at (1.1,1.1) {};
\vertex [crossdot] (e) at (2.6,1.1) {};
\vertex [crossdot] (f) at (3,0) {};
\propag [fer, revmom'={$p$}] (b) to (a);
\propag [fer] (c) to (b);
\propag [fer, mom'={$q$}] (d) to (b);
\propag [fer, mom'={$r$}] (e) to (c);
\propag [fer, mom={$-p-q-r$}] (f) to (c);
\end{feynhand}
\end{tikzpicture}
\notag\\
   &=\frac{1}{2!}\int_0^tds\,\int_0^sdu\,\int_{q,r}
   K_{t-s}(p)_{\mu\alpha}
   X^{(2,0)}(p,q,-p-q)_{\alpha\beta\eta}^{abe}
\notag\\
   &\qquad{}
   \times K_{s-u}(p+q)_{\eta\xi}
   X^{(2,0)}(p+q,r,-p-q-r)_{\xi\gamma\delta}^{ecd}
\notag\\
   &\qquad{}
   \times K_s(q)_{\beta\nu}A_\nu^{\prime b}(-q)K_u(r)_{\gamma\rho}A_\rho^{\prime c}(-r)
   K_u(p+q+r)_{\delta\sigma}A_\rho^{\prime d}(p+q+r).
\label{eq:(A10)}
\end{align}

For the FP ghost, the flow equation~\eqref{eq:(2.7)} can be expressed by the
integral equation,
\begin{equation}
   c^{\prime a}(t,x)
   =\int d^Dy\left[
   K_t(x-y)c^{\prime a}(y)
   +\int_0^tds\,K_{t-s}(x-y)
   \alpha_0f^{abc}A_\mu^{\prime b}(s,y)\partial_\mu c^{\prime c}(s,y)
   \right],
\label{eq:(A11)}
\end{equation}
where the heat kernel~$K_t(x)$ is given by
\begin{equation}
   K_t(x):=\int_pe^{ipx}e^{-\alpha_0tp^2},\qquad
   K_t(p):=e^{-\alpha_0tp^2}
\label{eq:(A12)}
\end{equation}
The flow vertex in~Eq.~\eqref{eq:(A11)} is represented as
\begin{align}
&\begin{tikzpicture}[baseline=(a.base)]
\begin{feynhand}
\vertex [ringblob,minimum height=3.5mm,minimum width=3.5mm] (a) at (0,0) {};
\vertex (b) at (-2,0) {$(q,b)$};
\vertex (c) at (1.4,1.4) {$(p,\mu,a)$};
\vertex (d) at (2,0) {$(r,c)$};
\propag [chasca, revmom'={$q$}] (a) to (b);
\propag [fer, mom'={$p$}] (c) to (a);
\propag [chasca, mom={$r$}] (d) to (a);
\end{feynhand}
\end{tikzpicture}
   =X^{(1,1)}(p,q,r)_\mu^{abc}
\notag\\
   &=\alpha_0if^{abc}r_\mu.
\label{eq:(A13)}
\end{align}
The expansion for the flowed ghost field~$c^{\prime a}(t,p)$ is thus given by
\begin{equation}
\begin{tikzpicture}
\begin{feynhand}
\vertex [squaredot] (a) at (0,0) {};
\vertex [crossdot] (b) at (2,0) {};
\propag [chasca, revmom'={$p$}] (b) to (a);
\end{feynhand}
\end{tikzpicture}
   =K_t(p)c^{\prime a}(p),
\label{eq:(A14)}
\end{equation}
\begin{align}
&\begin{tikzpicture}[baseline=(a.base)]
\begin{feynhand}
\vertex [ringblob,minimum height=3.5mm,minimum width=3.5mm] (a) at (0,0) {};
\vertex [squaredot] (b) at (-1.5,0) {};
\vertex [crossdot] (c) at (1.1,1.1) {};
\vertex [crossdot] (d) at (1.5,0) {};
\propag [chasca, revmom'={$p$}] (a) to (b);
\propag [fer, mom'={$q$}] (c) to (a);
\propag [chasca, mom={$-p-q$}] (d) to (a);
\end{feynhand}
\end{tikzpicture}
\notag\\
   &=\int_0^tds\,\int_{q,r}K_{t-s}(p)
   X^{(1,1)}(q,p,-p-q)_\alpha^{bac}
   K_s(q)_{\alpha\mu}A_\mu^{\prime b}(-q)K_s(p+q)c^{\prime c}(p+q),
\label{eq:(A15)}
\end{align}
and
\begin{align}
&\begin{tikzpicture}[baseline=(b.base)]
\begin{feynhand}
\vertex [squaredot] (a) at (-1.5,0) {};
\vertex [ringblob,minimum height=3.5mm,minimum width=3.5mm] (b) at (0,0) {};
\vertex [ringblob,minimum height=3.5mm,minimum width=3.5mm] (c) at (1.5,0) {};
\vertex [crossdot] (d) at (1.1,1.1) {};
\vertex [crossdot] (e) at (2.6,1.1) {};
\vertex [crossdot] (f) at (3,0) {};
\propag [chasca, revmom'={$p$}] (b) to (a);
\propag [chasca] (c) to (b);
\propag [fer, mom'={$q$}] (d) to (b);
\propag [fer, mom'={$r$}] (e) to (c);
\propag [chasca, mom={$-p-q-r$}] (f) to (c);
\end{feynhand}
\end{tikzpicture}
\notag\\
   &=\int_0^tds\,\int_0^sdu\,\int_{q,r}
   K_{t-s}(p)
   X^{(1,1)}(q,p,-p-q)_\alpha^{bae}
\notag\\
   &\qquad{}
   \times K_{s-u}(p+q)
   X^{(1,1)}(r,p+q,-p-q-r)_\beta^{ced}
\notag\\
   &\qquad{}
   \times K_s(q)_{\alpha\mu}A_\mu^{\prime b}(-q)K_u(r)_{\beta\nu}A_\nu^{\prime c}(-r)
   K_u(p+q+r)c^{\prime d}(p+q+r).
\label{eq:(A16)}
\end{align}

\subsection{Quantum propagators}
\label{sec:A.2}
So far, we have solved the flow equations~\eqref{eq:(2.5)} and~\eqref{eq:(2.7)}
perturbatively. The initial values for the flow equations, $A_\mu'(x)$
and~$c'(x)$, which are represented by the cross in the above diagrams, are
the subject of the functional integral in~Eq.~\eqref{eq:(2.2)}. We thus need
the propagators among~$A_\mu'(x)$, $c'(x)$ and~$\Bar{c}'(x)$.

For this, we note that besides the kinetic terms in the bare
action~\eqref{eq:(2.4)}, the Gaussian factors effectively produce ``mass
terms'' for fields as
\begin{align}
   &-\frac{\Lambda^2}{2}
   \int_p\frac{1}{\Lambda^{2\epsilon}g_\Lambda^2}
   A_\mu^{\prime a}(t,-p)A_\mu^{\prime a}(t,p)
\notag\\
   &=-\frac{1}{2g_0^2}\int_pZ\Lambda^2K_t(p)_{\mu\alpha}
   A_\alpha^{\prime a}(-p)K_t(p)_{\mu\beta}A_\beta^{\prime a}(p)
   +O(A^{\prime3})
\notag\\
   &=-\frac{1}{2g_0^2}\int_pZ\Lambda^2
   A_\mu^{\prime a}(-p)
   \left[
   \left(\delta_{\mu\nu}-\frac{p_\mu p_\nu}{p^2}\right)
   Z\Lambda^2e^{-2tp^2}
   +\frac{p_\mu p_\nu}{p^2}Z\Lambda^2e^{-2\alpha_0tp^2}
   \right]
   A_\nu^{\prime a}(p)
   +O(A^{\prime3}).
\label{eq:(A17)}
\end{align}
Combined with the kinetic operator in the Yang--Mills action, the tree-level
gauge propagator is given by
\begin{equation}
\begin{tikzpicture}
\begin{feynhand}
\vertex (a) at (-1,0) {};
\vertex (b) at (1,0) {};
\propag [glu] (b) to (a);
\end{feynhand}
\end{tikzpicture}
   :=\left\langle A_\mu^{\prime a}(p)A_\nu^{\prime b}(q)\right\rangle^{\text{tree}}
   =(2\pi)^D\delta(p+q)\delta^{ab}g_0^2D(p)_{\mu\nu},
\label{eq:(A18)}
\end{equation}
where
\begin{equation}
   D(p)_{\mu\nu}
   :=\left(\delta_{\mu\nu}-\frac{p_\mu p_\nu}{p^2}\right)
   \frac{1}{p^2+Z\Lambda^2e^{-2tp^2}}
   +\frac{p_\mu p_\nu}{p^2}
   \frac{\xi_0}{p^2+\xi_0Z\Lambda^2e^{-2\alpha_0tp^2}}.
\label{eq:(A19)}
\end{equation}
This curly line connects crosses in the flow line diagram. Then, it is
convenient to introduce $D_{s+u}(p)_{\mu\nu}$ as the product of the propagator
and heat kernels as~\cite{Luscher:2010iy,Luscher:2011bx}
\begin{align}
&\begin{tikzpicture}[baseline=(a.base)]
\begin{feynhand}
\vertex [squaredot] (a) at (-3,0) {};
\vertex [crossdot] (b) at (-1,0) {};
\vertex [crossdot] (c) at (1,0) {};
\vertex [squaredot] (d) at (3,0) {};
\propag [fer] (b) to (a);
\propag [glu] (c) to (b);
\propag [fer] (c) to (d);
\end{feynhand}
\end{tikzpicture}
\notag\\
&:=
\begin{tikzpicture}
\begin{feynhand}
\vertex (a) at (-1,0) {};
\vertex (b) at (1,0) {};
\propag [glu] (b) to (a);
\end{feynhand}
\end{tikzpicture}
\notag\\
   &:=\left\langle A_\mu^{\prime a}(s,p)A_\nu^{\prime b}(u,q)\right\rangle^{\text{tree}}
   =(2\pi)^D\delta(p+q)\delta^{ab}g_0^2
   D_{s+u}(p)_{\mu\nu},
\label{eq:(A20)}
\end{align}
where
\begin{align}
   D_{s+u}(p)_{\mu\nu}
   &:=K_s(p)_{\mu\rho}D(p)_{\rho\sigma}K_u(p)_{\sigma\nu}
\notag\\
   &=\left(\delta_{\mu\nu}-\frac{p_\mu p_\nu}{p^2}\right)
   \frac{e^{-(s+u)p^2}}{p^2+Z\Lambda^2e^{-2tp^2}}
   +\frac{p_\mu p_\nu}{p^2}
   \frac{\xi_0e^{-\alpha_0(s+u)p^2}}{p^2+\xi_0Z\Lambda^2e^{-2\alpha_0tp^2}}.
\label{eq:(A21)}
\end{align}

Similarly, for the ghost field,
\begin{equation}
\begin{tikzpicture}
\begin{feynhand}
\vertex (a) at (-1,0) {};
\vertex (b) at (1,0) {};
\propag [gho] (b) to (a);
\end{feynhand}
\end{tikzpicture}
   :=\left\langle c^{\prime a}(p)\Bar{c}^{\prime b}(q)\right\rangle^{\text{tree}}
   =(2\pi)^D\delta(p+q)\delta^{ab}g_0^2D(p),
\label{eq:(A22)}
\end{equation}
where
\begin{equation}
   D(p):=\frac{1}{p^2+Z_{\Bar{c}}Z\Lambda^2e^{-\alpha_0tp^2}},
\label{eq:(A23)}
\end{equation}
and
\begin{align}
&\begin{tikzpicture}[baseline=(a.base)]
\begin{feynhand}
\vertex [squaredot] (a) at (-3,0) {};
\vertex [crossdot] (b) at (-1,0) {};
\vertex (c) at (1,0) {};
\propag [chasca] (b) to (a);
\propag [gho] (c) to (b);
\end{feynhand}
\end{tikzpicture}
\notag\\
&:=
\begin{tikzpicture}
\begin{feynhand}
\vertex (a) at (-1,0) {};
\vertex (b) at (1,0) {};
\propag [gho] (b) to (a);
\end{feynhand}
\end{tikzpicture}
\notag\\
   &:=\left\langle c^{\prime a}(s,p)\Bar{c}^{\prime b}(q)\right\rangle^{\text{tree}}
   =(2\pi)^D\delta(p+q)\delta^{ab}g_0^2D_s(p),
\label{eq:(A24)}
\end{align}
where
\begin{equation}
   D_s(p)
   :=K_s(p)D(p)
   =\frac{e^{-\alpha_0sp^2}}{p^2+Z_{\Bar{c}}Z\Lambda^2e^{-\alpha_0tp^2}}.
\label{eq:(A25)}
\end{equation}

\subsection{Vertices}
\label{sec:A.3}
The interaction vertices arising from the bare action~\eqref{eq:(2.4)} are
conventional ones. In momentum space:
\begin{align}
&\begin{tikzpicture}[baseline=(a.base)]
\begin{feynhand}
\vertex [dot] (a) at (0,0) {};
\vertex (b) at (-1.8,0) {$(p,\mu,a)$};
\vertex (c) at (1.1,1.1) {$(q,\nu,b)$};
\vertex (d) at (1.1,-1.1) {$(r,\rho,c)$};
\propag [glu, revmom'={$p$}] (a) to (b);
\propag [glu, mom'={$q$}] (c) to (a);
\propag [glu, mom={$r$}] (d) to (a);
\end{feynhand}
\end{tikzpicture}
\notag\\
   &=\frac{1}{3!}\frac{i}{g_0^2}f^{abc}
   \left[
   (r-q)_\mu\delta_{\nu\rho}
   +(p-r)_\nu\delta_{\rho\mu}
   +(q-p)_\rho\delta_{\mu\nu}
   \right],
\label{eq:(A26)}
\end{align}
and
\begin{align}
&\begin{tikzpicture}[baseline=(a.base)]
\begin{feynhand}
\vertex [dot] (a) at (0,0) {};
\vertex (b) at (-1.1,1.1) {$(p,\mu,a)$};
\vertex (c) at (1.1,1.1) {$(q,\nu,b)$};
\vertex (d) at (1.1,-1.1) {$(r,\rho,c)$};
\vertex (e) at (-1.1,-1.1) {$(s,\sigma,d)$};
\propag [glu, revmom'={$p$}] (a) to (b);
\propag [glu, mom={$q$}] (c) to (a);
\propag [glu, mom={$r$}] (d) to (a);
\propag [glu, revmom'={$s$}] (a) to (e);
\end{feynhand}
\end{tikzpicture}
\notag\\
   &=-\frac{1}{4!g_0^2}\Bigl[
   f^{abe}f^{cde}\left(
   \delta_{\mu\rho}\delta_{\nu\sigma}-\delta_{\mu\sigma}\delta_{\nu\rho}
   \right)
   +f^{ace}f^{dbe}\left(
   \delta_{\mu\sigma}\delta_{\rho\nu}-\delta_{\mu\nu}\delta_{\rho\sigma}
   \right)
\notag\\
   &\qquad\qquad\qquad{}
   +f^{ade}f^{bce}\left(
   \delta_{\mu\nu}\delta_{\sigma\rho}-\delta_{\mu\rho}\delta_{\sigma\nu}
   \right)\Bigr],
\label{eq:(A27)}
\end{align}
and
\begin{equation}
\begin{tikzpicture}[baseline=(b.base)]
\begin{feynhand}
\vertex [dot] (a) at (0,0) {};
\vertex (b) at (-1.5,0) {$(p,a)$};
\vertex (c) at (0,1.3) {$(q,\mu,b)$};
\vertex (d) at (1.5,0) {$(r,c)$};
\propag [gho, mom'={$p$}] (b) to (a);
\propag [glu, revmom'={$q$}] (a) to (c);
\propag [gho, mom={$r$}] (d) to (a);
\end{feynhand}
\end{tikzpicture}
   =\frac{1}{g_0^2}if^{abc}p_\mu,
\label{eq:(A28)}
\end{equation}
where $(r,c)$ is the vertex for the ghost field~$c^{\prime c}(-r)$.

In the present system, we have additional vertices arising from the Gaussian
factors in~Eq.~\eqref{eq:(2.2)}, though the perturbative solutions for
$A_\mu'(t,x)$ and~$c'(t,x)$; these are given by substituting
Eqs.~\eqref{eq:(A1)} and~\eqref{eq:(A11)} into the Gaussian factors.
Diagrammatically, we represent those interaction vertices by a shaded blob.
In lower orders, we have
\begin{align}
&
\begin{tikzpicture}[baseline=(b.base)]
\begin{feynhand}
\vertex [crossdot] (a) at (-1.5,0) {};
\vertex [grayblob,minimum height=3.5mm,minimum width=3.5mm] (b) at (0,0) {};
\vertex [crossdot] (c) at (1.1,1.1) {};
\vertex [crossdot] (d) at (1.1,-1.1) {};
\propag [fer, revmom'={$p$}] (b) to (a);
\propag [fer, mom'={$q$}] (c) to (b);
\propag [fer, mom={$r$}] (d) to (b);
\end{feynhand}
\end{tikzpicture}
\notag\\
&:=
\begin{tikzpicture}[baseline=(b.base)]
\begin{feynhand}
\vertex [crossdot] (x) at (-3,0) {};
\vertex [squaredot] (a) at (-1.5,0) {};
\vertex [ringblob,minimum height=3.5mm,minimum width=3.5mm] (b) at (0,0) {};
\vertex [crossdot] (c) at (1.1,1.1) {};
\vertex [crossdot] (d) at (1.1,-1.1) {};
\propag [fer, mom={$p$}] (x) to (a);
\propag [fer, revmom'={$p$}] (b) to (a);
\propag [fer, mom'={$q$}] (c) to (b);
\propag [fer, mom={$r$}] (d) to (b);
\end{feynhand}
\end{tikzpicture}
+\text{permutations}
\notag\\
   &=-\frac{1}{g_0^2}Z\Lambda^2A_\mu^{\prime a}(-p)
\notag\\
   &\qquad{}
   \times
   \frac{1}{2!}\int_0^tds\,\int_qK_{2t-s}(p)_{\mu\alpha}
   X^{(2,0)}(p,q,r)_{\alpha\beta\gamma}^{abc}
   K_s(q)_{\beta\nu}A_\nu^{\prime b}(-q)K_s(r)_{\gamma\rho}A_\rho^{\prime c}(-r),
\label{eq:(A29)}
\end{align}
and
\begin{align}
&\begin{tikzpicture}[baseline=(a.base)]
\begin{feynhand}
\vertex [grayblob,minimum height=3.5mm,minimum width=3.5mm] (a) at (0,0) {};
\vertex [crossdot] (b) at (-1.1,1.1) {};
\vertex [crossdot] (c) at (1.1,1.1) {};
\vertex [crossdot] (d) at (1.1,-1.1) {};
\vertex [crossdot] (e) at (-1.1,-1.1) {};
\propag [fer, mom'={$p$}] (b) to (a);
\propag [fer, mom'={$q$}] (c) to (a);
\propag [fer, mom'={$r$}] (d) to (a);
\propag [fer, mom'={$s$}] (e) to (a);
\end{feynhand}
\end{tikzpicture}
\notag\\
&:=
\begin{tikzpicture}[baseline=(x.base)]
\begin{feynhand}
\vertex [crossdot] (y) at (-4.1,1.1) {};
\vertex [crossdot] (z) at (-4.1,-1.1) {};
\vertex [ringblob,minimum height=3.5mm,minimum width=3.5mm] (x) at (-3,0) {};
\vertex [squaredot] (a) at (-1.5,0) {};
\vertex [ringblob,minimum height=3.5mm,minimum width=3.5mm] (b) at (0,0) {};
\vertex [crossdot] (c) at (1.1,1.1) {};
\vertex [crossdot] (d) at (1.1,-1.1) {};
\propag [fer] (x) to (a);
\propag [fer] (b) to (a);
\propag [fer, mom'={$q$}] (c) to (b);
\propag [fer, mom={$r$}] (d) to (b);
\propag [fer, mom={$p$}] (y) to (x);
\propag [fer, mom'={$s$}] (z) to (x);
\end{feynhand}
\end{tikzpicture}
\notag\\
&\qquad{}
+\begin{tikzpicture}[baseline=(b.base)]
\begin{feynhand}
\vertex [crossdot] (x) at (-3,0) {};
\vertex [squaredot] (a) at (-1.5,0) {};
\vertex [ringblob,minimum height=3.5mm,minimum width=3.5mm] (b) at (0,0) {};
\vertex [crossdot] (c) at (1.1,1.1) {};
\vertex [crossdot] (d) at (1.5,0) {};
\vertex [crossdot] (e) at (1.1,-1.1) {};
\propag [fer, mom={$p$}] (x) to (a);
\propag [fer] (b) to (a);
\propag [fer, mom'={$q$}] (c) to (b);
\propag [fer, mom'={$r$}] (d) to (b);
\propag [fer, mom={$s$}] (e) to (b);
\end{feynhand}
\end{tikzpicture}
\notag\\
&\qquad{}
+\begin{tikzpicture}[baseline=(b.base)]
\begin{feynhand}
\vertex [crossdot] (x) at (-3,0) {};
\vertex [squaredot] (a) at (-1.5,0) {};
\vertex [ringblob,minimum height=3.5mm,minimum width=3.5mm] (b) at (0,0) {};
\vertex [ringblob,minimum height=3.5mm,minimum width=3.5mm] (c) at (1.5,0) {};
\vertex [crossdot] (d) at (1.1,1.1) {};
\vertex [crossdot] (e) at (2.6,1.1) {};
\vertex [crossdot] (f) at (3,0) {};
\propag [fer, mom={$p$}] (x) to (a);
\propag [fer] (b) to (a);
\propag [fer] (c) to (b);
\propag [fer, mom'={$q$}] (d) to (b);
\propag [fer, mom'={$r$}] (e) to (c);
\propag [fer, mom={$s$}] (f) to (c);
\end{feynhand}
\end{tikzpicture}
\notag\\
&\qquad{}
+\text{permutations}.
\label{eq:(A30)}
\end{align}

Similarly, for the FP ghost fields,
\begin{align}
&\begin{tikzpicture}[baseline=(b.base)]
\begin{feynhand}
\vertex [crossdot] (a) at (-1.5,0) {};
\vertex [grayblob,minimum height=3.5mm,minimum width=3.5mm] (b) at (0,0) {};
\vertex [crossdot] (c) at (1.1,1.1) {};
\vertex [crossdot] (d) at (1.5,0) {};
\propag [gho, revmom'={$p$}] (b) to (a);
\propag [fer, mom'={$q$}] (c) to (b);
\propag [chasca, mom={$r$}] (d) to (b);
\end{feynhand}
\end{tikzpicture}
\notag\\
&:=\begin{tikzpicture}[baseline=(b.base)]
\begin{feynhand}
\vertex [crossdot] (x) at (-3,0) {};
\vertex [squaredot] (a) at (-1.5,0) {};
\vertex [ringblob,minimum height=3.5mm,minimum width=3.5mm] (b) at (0,0) {};
\vertex [crossdot] (c) at (1.1,1.1) {};
\vertex [crossdot] (d) at (1.5,0) {};
\propag [chasca] (b) to (a);
\propag [gho, mom={$p$}] (x) to (a);
\propag [fer, mom'={$q$}] (c) to (b);
\propag [chasca, mom={$r$}] (d) to (b);
\end{feynhand}
\end{tikzpicture}
\notag\\
&=-\frac{1}{g_0^2}Z_{\Bar{c}}Z\Lambda^2\Bar{c}^{\prime a}(-p)
   \int_0^tds\,\int_{q,r}K_{t-s}(p)
   X^{(1,1)}(q,p,r)_\alpha^{bac}
   K_s(q)_{\alpha\mu}A_\mu^{\prime b}(-q)K_s(r)c^{\prime c}(-r),
\label{eq:(A31)}
\end{align}
and
\begin{align}
&\begin{tikzpicture}[baseline=(b.base)]
\begin{feynhand}
\vertex [crossdot] (a) at (-1.5,0) {};
\vertex [grayblob,minimum height=3.5mm,minimum width=3.5mm] (b) at (0,0) {};
\vertex [crossdot] (c) at (-1.1,1.1) {};
\vertex [crossdot] (d) at (1.1,1.1) {};
\vertex [crossdot] (e) at (1.5,0) {};
\propag [gho, mom'={$p$}] (a) to (b);
\propag [fer, mom={$q$}] (c) to (b);
\propag [fer, mom'={$r$}] (d) to (b);
\propag [chasca, mom={$s$}] (e) to (b);
\end{feynhand}
\end{tikzpicture}
\notag\\
&=\begin{tikzpicture}[baseline=(b.base)]
\begin{feynhand}
\vertex [crossdot] (x) at (-3,0) {};
\vertex [squaredot] (a) at (-1.5,0) {};
\vertex [ringblob,minimum height=3.5mm,minimum width=3.5mm] (b) at (0,0) {};
\vertex [ringblob,minimum height=3.5mm,minimum width=3.5mm] (c) at (1.5,0) {};
\vertex [crossdot] (d) at (1.1,1.1) {};
\vertex [crossdot] (e) at (2.6,1.1) {};
\vertex [crossdot] (f) at (3,0) {};
\propag [chasca] (b) to (a);
\propag [chasca] (c) to (b);
\propag [gho, mom'={$p$}] (x) to (a);
\propag [fer, mom'={$q$}] (d) to (b);
\propag [fer, mom'={$r$}] (e) to (c);
\propag [chasca, mom={$s$}] (f) to (c);
\end{feynhand}
\end{tikzpicture}
+\text{permutations}.
\label{eq:(A32)}
\end{align}

Diagrams in Figs~\ref{fig:1} and~\ref{fig:2} in the text are drawn following
to the above Feynman rule.

\section{Calculation of diagrams in~Fig.~\ref{fig:1}}
\label{sec:B}
In this appendix, we present the calculation of one-loop diagrams
in~Fig.~\ref{fig:1}, except for the diagram~(d) for which the calculation is
given in the main text. Notation given in~Appendix~\ref{sec:A} is understood.
Since for~Eq.~\eqref{eq:(2.27)} we have to find
$\lim_{|\ell|\to\infty}\left.\mathcal{I}(\ell,p)\right|_{O(p^2)}$,
where $\mathcal{I}(\ell,p)$ is the integrand of the loop integral, we may
appropriately set $p\to0$ in expressions as long as
$\lim_{|\ell|\to\infty}\left.\mathcal{I}(\ell,p)\right|_{O(p^2)}$ is not changed.
In what follows, such a simplification step is denoted by the arrow ($\to$).

\begin{align}
   &G_{2t}(p)_{\mu\nu}^{ab}[\text{diagram~(e) of~Fig.~\ref{fig:1}}]
\notag\\
   &=\frac{1}{2}g_0^4\int_0^tds\,\int_\ell\,
   K_{t-s}(p)_{\mu\alpha}
   X^{(2,0)}(p,\ell,-\ell-p)_{\alpha\beta\gamma}^{acd}
\notag\\
   &\qquad{}
   \times
   \int_0^tdu\,K_{t-u}(p)_{\nu\delta}
   X^{(2,0)}(-p,-\ell,p+\ell)_{\alpha\beta\gamma}^{bcd}
   D_{s+u}(\ell)_{\beta\epsilon}D_{s+u}(\ell+p)_{\gamma\tau}.
\label{eq:(B1)}
\end{align}
By examining the integrand after the integrations over $s$ and~$u$, we find
that the integrand behaves~$\sim1/(\ell^2)^3$ for~$|\ell|\to\infty$. Therefore,
this diagram does not contribute to the limit in~Eq.~\eqref{eq:(2.27)}.

\begin{align}
   &G_{2t}(p)_{\mu\nu}^{ab}[\text{diagram~(f) of~Fig.~\ref{fig:1}}]
\notag\\
   &=g_0^4\int_0^tds\,K_{t-s}(p)_{\mu\alpha}\int_\ell\,
   X^{(3,0)}(p,\ell,-\ell,-p)_{\alpha\beta\gamma\delta}^{accb}
   D_{2s}(\ell)_{\beta\gamma}D_{s+t}(p)_{\delta\nu}
\notag\\
   &\to
   g_0^4\int_0^tds\,\int_\ell\,
   X^{(3,0)}(0,\ell,-\ell,0)_{\mu\beta\gamma\delta}^{accb}
   D_{2s}(\ell)_{\beta\gamma}D_{s+t}(p)_{\delta\nu}
\notag\\
   &=2g_0^4C_A\delta^{ab}\int_0^t ds\,\int_\ell\,
   \left[
   \left(2-\frac{1}{D}-D\right)\frac{e^{-2s\ell^2}}{\ell^2}
   +\left(\frac{1}{D}-1\right)\xi_0\frac{e^{-2\alpha_0s\ell^2}}{\ell^2}
   \right]
   D_t(p)_{\mu\nu}
\notag\\
   &\to
   g_\Lambda^4C_A\delta^{ab}\times
   (-1)\left(1-\frac{1}{D}\right)\left(D-1+\frac{\xi_\Lambda}{\alpha_0}\right)
   \int_\ell\,\frac{1}{(\ell^2)^2}D_t(p)_{\mu\nu}.
\label{eq:(B2)}
\end{align}

\begin{align}
   &G_{2t}(p)_{\mu\nu}^{ab}[\text{diagram~(g) of~Fig.~\ref{fig:1}}]
\notag\\
   &=2g_0^4\int_0^tds\,\int_0^sdu\,K_{t-s}(p)_{\mu\alpha}\int_\ell\,
   X^{(2,0)}(p,\ell,-\ell-p)_{\alpha\beta\gamma}^{acd}
   D_{s+u}(\ell)_{\beta\sigma}K_{s-u}(\ell+p)_{\gamma\tau}
\notag\\
   &\qquad\qquad\qquad{}
   \times
   X^{(2,0)}(\ell+p,-\ell,-p)_{\tau\sigma\rho}^{dcb}
   D_{u+t}(p)_{\rho\nu}
\notag\\
   &\to
   2g_0^4\int_0^tds\,\int_0^sdu\,K_t(p)_{\mu\alpha}\int_\ell\,
   X^{(2,0)}(0,\ell,-\ell)_{\alpha\beta\gamma}^{acd}
   D_{s+u}(\ell)_{\beta\sigma}K_{s-u}(\ell)_{\gamma\tau}
\notag\\
   &\qquad\qquad\qquad{}
   \times
   X^{(2,0)}(\ell,-\ell,0)_{\tau\sigma\rho}^{dcb}
   D_t(p)_{\rho\nu}
\notag\\
   &=2g_0^4C_A\delta^{ab}K_t(p)_{\mu\alpha}
   \int_0^tds\,\int_0^sdu\,\int_\ell\,
\notag\\
   &\qquad{}
   \times
   \biggl(\frac{\ell_\alpha\ell_\rho}{\ell^2}
   \left\{4(D-1)e^{-2s\ell^2}
   +\left[-2-2(\alpha_0-1)(\alpha_0+1)\right]
   \xi_0e^{-2\alpha_0s\ell^2}
   \right\}
\notag\\
   &\qquad\qquad{}
   +\left(\delta_{\alpha\rho}-\frac{\ell_\alpha\ell_\rho}{\ell^2}\right)
\notag\\
   &\qquad\qquad\qquad{}
   \times\left\{(\alpha_0+1)e^{-[(\alpha_0+1)s-(\alpha_0-1)u]\ell^2}
   -(\alpha_0-1)(\alpha_0+1)\xi_0e^{-[(\alpha_0+1)s+(\alpha_0-1)u]\ell^2}
   \right\}\biggr)
\notag\\
   &\qquad\qquad\qquad\qquad{}
   \times D_t(p)_{\rho\nu}
\notag\\
   &\to
   g_\Lambda^4C_A\delta^{ab}\times
   \left\{
   3\left(1-\frac{1}{D}\right)
   +\left[-\frac{1}{D}-\left(1-\frac{1}{D}\right)\left(1-\frac{1}{\alpha_0}
   \right)\right]\xi_\Lambda
   \right\}
   \int_\ell\,\frac{1}{(\ell^2)^2}
   D_t(p)_{\mu\nu}.
\label{eq:(B3)}
\end{align}

The large loop momentum behaviors of the integrand for the diagrams (a), (b),
and~(c) are identical to those for the usual Yang--Mills theory because in
these diagrams all vertices are on the flow time~$0$ (i.e., no
integration over the flow time).

\begin{align}
   &G_{2t}(p)_{\mu\nu}^{ab}[\text{diagram~(a) of~Fig.~\ref{fig:1}}]
\notag\\
   &=\frac{1}{2}g_0^4C_A
   D(p)_{\mu\alpha}D(p)_{\rho\nu}
\notag\\
   &\qquad{}
   \times
   \int d\ell\,
   \left[
   (2\ell+p)_\alpha\delta_{\beta\gamma}
   -(\ell+2p)_\beta\delta_{\gamma\alpha}
   -(\ell-p)_\gamma\delta_{\alpha\beta}
   \right]
\notag\\
   &\qquad\qquad\qquad{}
   \times
   \left[
   (2\ell+p)_\rho\delta_{\sigma\tau}
   -(\ell+2p)_\sigma\delta_{\tau\rho}
   -(\ell-p)_\tau\delta_{\rho\sigma}
   \right]
   D(\ell)_{\beta\sigma}D(\ell+p)_{\gamma\tau}.
\label{eq:(B4)}
\end{align}
After a lengthy calculation, we have
\begin{align}
   &G_{2t}(p)_{\mu\nu}^{ab}[\text{diagram~(a) of~Fig.~\ref{fig:1}}]
\notag\\
   &\to
   g_\Lambda^4C_A\delta^{ab}\times(-1)
   \int_\ell\,
   \frac{1}{(\ell^2)^2}
   \left[
   \left(\frac{25}{12}-\frac{1}{2}\xi_\Lambda\right)
   \left(\delta_{\mu\nu}p^2-p_\mu p_\nu\right)
   -\frac{1}{4}\xi_Ap_\mu p_\nu
   \right].
\label{eq:(B5)}
\end{align}

Diagram~(b) of~Fig.~\ref{fig:1} does not contribute to~Eq.~\eqref{eq:(2.27)}
because the loop integral contains only one propagator and the power counting
shows that it cannot produce an $O(p^2)$ term
in~$\mathcal{I}(\ell,p)_{\mu\nu}^{ab}$. Alternatively, by invoking dimensional
regularization, the loop integral vanishes.

\begin{align}
   &G_{2t}(p)_{\mu\nu}^{ab}[\text{diagram~(c) of~Fig.~\ref{fig:1}}]
\notag\\
   &=-g_0^4C_A\delta^{ab}D_{\mu\alpha}(p)
   \int_\ell\,\ell_\alpha(\ell+p)_\beta D(\ell)D(\ell+p)D_{\beta\nu}(p)
\notag\\
   &\to
   -g_0^4C_A\delta^{ab}D_{\mu\alpha}(p)
   \int_\ell\,\ell_\alpha(\ell+p)_\beta
   \frac{1}{\ell^2}\frac{1}{(\ell+p)^2}D_{\beta\nu}(p)
\notag\\
   &\to
   g_\Lambda^4C_A\delta^{ab}\times(-1)
   \int_\ell\,
   \frac{1}{(\ell^2)^2}
   \left[
   \frac{1}{12}\left(\delta_{\mu\nu}p^2-p_\mu p_\nu\right)
   +\frac{1}{4}\xi_\Lambda p_\mu p_\nu
   \right].
\label{eq:(B6)}
\end{align}



%



\let\doi\relax










\end{document}